\documentclass[aps,prd,preprint,groupedaddress,preprintnumbers,longbibliography,nofootinbib]{revtex4-1}

\usepackage{amsmath}
\usepackage{amsfonts}
\usepackage{graphicx}

\newcommand{\nt}{\notag}
\newcommand{\ep}{\epsilon}
\newcommand{\vev}[1]{\left\langle #1 \right\rangle}
\newcommand{\vvev}[1]{\vev{\kern-0.25em\vev{#1}\kern-0.25em}}
\newcommand{\N}{\mathcal{N}}
\newcommand{\Op}{\mathcal{O}}
\newcommand{\op}[1]{\left[ #1 \right]}
\newcommand{\PP}{\mathcal{P}}
\newcommand{\lb}{\left\lbrace}
\newcommand{\rb}{\right\rbrace}

\allowdisplaybreaks

\begin{document}

\preprint{KOBE-TH-20-05}

\title{The Operator Algebra at the Gaussian Fixed-Point}


\author{H.~Sonoda}
\email[]{hsonoda@kobe-u.ac.jp}
\affiliation{Physics Department, Kobe University, Kobe 657-8501, JAPAN}


\date{\today}

\begin{abstract}
  We consider the multiple products of relevant and marginal scalar
  composite operators at the Gaussian fixed-point in $D=4$ dimensions.
  This amounts to perturbative construction of the $\phi^4$ theory
  where the parameters of the theory are momentum dependent sources.
  Using the exact renormalization group (ERG) formalism, we show how
  the scaling properties of the sources are given by the
  short-distance singularities of the multiple products.
\end{abstract}


\maketitle

\section{Introduction}

For a quantum field theory to have a well defined continuum limit, it
must have an ultraviolet (UV) fixed-point of the renormalization group.
Conversely, given a fixed-point, we can build a continuum limit with
as many parameters as the number of relevant operators.  Modulo the
important question of convergence, the continuum limit with parameters
can be constructed out of the multiple products of the relevant
operators at the fixed-point. \cite{Wilson:1973jj}

Following this idea, Cardy has considered a generic fixed-point action
$S^*$ with relevant operators $\Op_i$. \cite{Cardy:1996xt} Let the first
order action of the theory be
\begin{equation}
  S^* + \sum_i g_i \int d^D x\, \Op_i (x)\,.
\end{equation}
The short-distance singularity of the operator products are given 
as
\begin{equation}
  \Op_i (x) \Op_j (y) \overset{x \to y}{\longrightarrow} 
 \sum_k \frac{c_{ij,k}}{|x-y|^{x_i+x_j-x_k}} \Op_k (x)\,,
\end{equation}
where $x_i$ is the scale dimension of $\Op_i$ in $D$-dimensional
coordinate space.  When $x_i + x_j - x_k \ge D$, the product is
unintegrable, and the second order perturbation needs regularization.
Consequently, the parameter $g_i$ acquires a mixing term proportional
to $\sum_{j,k} c_{jk,i} g_j g_k$ under the renormalization group.
(This result has been reviewed in Appendix C of \cite{Pagani:2017gnd}
using the exact renormalization group formalism.)

Even before this observation, we all had been familiar with the idea
that the short-distance singularities are the source of
renormalization and the associated renormalization group.  Consider
the perturbative construction of $\phi^4$ theory in $D=4 - \ep$
dimensions.  The bare parameters and field are given by
\begin{subequations}
\begin{align}
  m_0^2 &= Z_m (\ep; \lambda) m^2\,,\\
  \lambda_0 &= Z_\lambda (\ep; \lambda) \lambda \mu^\ep\,,\\
  \phi_0 &= \sqrt{Z (\ep; \lambda)}\,\phi\,,
\end{align}
\end{subequations}
where the renormalization constants are given in the MS prescription.
The renormalization constants, which cancel the UV divergences, are
determined by the beta functions and the anomalous dimension as
\cite{tHooft:1973mfk}
\begin{subequations}\label{sec1-Z}
\begin{align}
  Z_\lambda (\ep; \lambda) &= \exp \left( \int_0^\lambda dx\,
                             \left(\frac{\ep}{\ep x + \beta (x)} -
                             \frac{1}{x} \right) \right) \,,\\
  Z_m (\ep; \lambda) &=  \exp \left( - \int_0^\lambda dx\,
                       \frac{\beta_m (x)}{\ep x + \beta (x)}\right)\,,\\
  Z (\ep; \lambda) &= \exp \left( - \int_0^\lambda dx \frac{2 \gamma
                     (x)}{\ep x + \beta (x)}\right)\,.
\end{align}
\end{subequations}
(Note that we have chosen the sign of beta functions to give
the change of parameters toward the IR.)

In this paper we take the Gaussian fixed-point in $D=4$ dimensions as
the simplest example, and study how to construct the multiple products
of three scalar composite operators
$\phi^2, \phi^4, \partial_\mu \phi \partial_\mu \phi$ which are either
relevant or marginal.  (These operators will be called
$\Op_2, \Op_4, \N$ in the main text.)  Using the exact renormalization
group (ERG) formalism, we construct a theory with momentum dependent
sources coupled to the three composite operators.  (There are many
reviews of ERG.  See \cite{Rosten:2010vm} and references therein.  We
mostly follow the conventions of \cite{Igarashi:2009tj} and
\cite{Sonoda:2015bla} out of familiarity.)  We will obtain a precise
relation between the short-distance singularities of the operator
products and the mixing coefficients under scaling of the sources.

Momentum dependent sources for composite operators, equivalent to
space dependent parameters, have been considered before.  In
generalizing Zamolodchikov's $c$-theorem, Jack and Osborn have
introduced space dependent parameters using the dimensional
regularization.  \cite{Jack:1990eb, Osborn:1991gm, Osborn:1991mk} (See
also \cite{Baume:2014rla} for more recent developments.)  The
correlation functions of the bare composite operators have poles in
$\ep = 4 - D$, which they have related to the beta functions of the
space dependent parameters, just as in (\ref{sec1-Z}). The poles
result from the short-distance singularities we discuss here.

This paper is organized as follows.  In Sec.~\ref{section-operators}
we briefly review the composite operators in the ERG formalism.  We
introduce the three scalar composite operators the multiple products
of which constitute the main subject of this paper.  In
Sec.~\ref{section-multiple} we sketch how to construct multiple
products of composite operators in the ERG formalism.  This section is
based upon the results of \cite{Pagani:2017tdr}.  In
Sec.~\ref{section-products-two} we give concrete examples of the
products of two composite operators.  We then give a general
discussion of the number operator in Sec.~\ref{section-number}.  The
number operator is an equation-of-motion operator, and the precise
control we gain over their products greatly simplifies our study of
the multiple products.  In Sec.~\ref{section-W} we introduce the
generating functional $W$ of the multiple products.  The sources for
$W$ are momentum dependent parameters of the $\phi^4$ theory.  We show
that the scaling property of $e^W$ is given by the coefficients of
short-distance singularities in the multiple products of composite
operators.  The paper is concluded in Sec.~\ref{section-conclusions}.
We sketch the calculations of some integrals of cutoff functions in
Appendix \ref{appendix:integrals}, derive the asymptotic behavior of
two functions ($F, G$) in Appendix \ref{appendix:H}, review quickly
the equation-of-motion operators in the ERG formalism in Appendix
\ref{appendix:eom}, and construct some products of three composite
operators in Appendix \ref{appendix:three}.  In Appendix
\ref{appendix:coordinates} we switch from momentum space to coordinate
space to discuss the scaling properties of the space dependent
parameters.

Throughout the paper we work in $D=4$ dimensional Euclidean space, and
use the following short-hand notation for the momentum space:
\[
  \int_p \equiv \int \frac{d^D p}{(2 \pi)^D},\quad
  \delta (p) \equiv (2 \pi)^D \delta^{(D)} (p),\quad
  p \cdot \partial_p \equiv \sum_{\mu=1}^D p_\mu
  \frac{\partial}{\partial p_\mu}\,.
\]

\section{Composite operators in the ERG formalism\label{section-operators}}

The Wilson action of the free massless scalar theory is given by
\begin{equation}
  S_\Lambda [\phi] = - \frac{1}{2} \int_p \phi (-p)
  \frac{p^2}{K(p/\Lambda)} \phi (p)\label{SLambda}
\end{equation}
in the momentum space, where $\Lambda$ is a momentum cutoff.  The
smooth and positive cutoff function $K(p/\Lambda)$ is $1$ at $p=0$, is
of order $1$ for $p \sim \Lambda$, and approaches zero rapidly as
$p/\Lambda \to \infty$.  An example is given by
$\exp (- p^2/\Lambda^2)$.  The corresponding propagator
$K(p/\Lambda)/p^2$ suppresses the high momentum modes.  The cutoff
dependence of the Wilson action is given by the ERG differential
equation\footnote{Strictly speaking, we must ignore the field
  independent part of the action for the validity of
  Eq.~(\ref{sec2-ERG-Lambda}).}
\begin{equation}
  - \Lambda \frac{\partial}{\partial \Lambda} e^{S_\Lambda [\phi]}
  = \int_p \frac{\Delta (p/\Lambda)}{p^2} \frac{1}{2} \frac{\delta^2}{\delta \phi (p)
      \delta \phi (-p)}\, e^{S_\Lambda [\phi]}\,,\label{sec2-ERG-Lambda}
\end{equation}
where we define
\begin{equation}
  \Delta (p/\Lambda) \equiv \Lambda \frac{\partial}{\partial \Lambda}
  K(p/\Lambda) > 0\,.
\end{equation}

To obtain the Gaussian fixed-point from $S_\Lambda$, we must rescale
the momentum and field using the cutoff $\Lambda$.  We introduce
dimensionless momentum $\bar{p}$ and field $\bar{\phi} (\bar{p})$ by
\begin{subequations}
\begin{align}
  p &= \Lambda \bar{p}\,,\\
  \phi (p) &= \Lambda^{- \frac{D+2}{2}} \bar{\phi} (\bar{p})\,.
\end{align}
\end{subequations}
Please note that the mass dimension of the scalar field $\phi (x)$ in
coordinate space is $\frac{D-2}{2}$, and that of its Fourier transform
$\phi (p) = \int d^D x \, e^{- i p x} \phi (x)$ is $- \frac{D+2}{2}$.
Writing (\ref{SLambda}) using the dimensionless momentum and field, we
obtain the Gaussian fixed-point action
\begin{equation}
  \bar{S} [\bar{\phi}] = - \frac{1}{2} \int_{\bar{p}} \bar{\phi}
  (-\bar{p}) \frac{\bar{p}^2}{K(\bar{p})} \bar{\phi} (\bar{p})\,.
\end{equation}
This satisfies the fixed-point equation
\begin{equation}
  0 = \int_{\bar{p}} \left[ \left( \frac{D+2}{2} + \bar{p} \cdot
    \partial_{\bar{p}} \right) \bar{\phi} (\bar{p}) \cdot
  \frac{\delta}{\delta \bar{\phi} (\bar{p})}
  + \frac{\Delta (\bar{p})}{\bar{p}^2} \frac{1}{2}
  \frac{\delta^2}{\delta \bar{\phi} (\bar{p}) \delta \bar{\phi}
    (-\bar{p})}\right]\, e^{\bar{S}
    [\bar{\phi}]}\,.
\end{equation}
From now on we use this dimensionless convention by measuring all the
dimensionful quantities in units of appropriate powers of the cutoff
$\Lambda$.  We omit the bars above momenta and fields entirely for the
sake of simplicity.  Hence, we write the Gaussian fixed-point action
as
\begin{equation}
  S [\phi] = - \frac{1}{2} \int_p \phi (-p) \frac{p^2}{K(p)} \phi
  (p)\,.
\end{equation}

The correlation functions defined by
\begin{align}
&  \vvev{\phi (p_1) \cdots \phi (p_n)}\nt\\
& \equiv \prod_{i=1}^n \frac{1}{K(p_i)}\cdot \vev{\exp \left( - \int_p
      \frac{K(p) \left(1 - K(p)\right)}{p^2} \frac{1}{2}
      \frac{\delta^2}{\delta \phi (p) \delta \phi (-p)}\right) \phi
    (p_1) \cdots \phi (p_n)}_S\label{sec2-def-corr}
\end{align}
satisfy the scaling relation
\begin{equation}
  \vvev{\phi (p_1 e^t) \cdots \phi (p_n e^t)}
  = e^{- n \frac{D+2}{2} t} \vvev{\phi (p_1) \cdots \phi (p_n)}\,.
\end{equation}
Here,
\begin{equation}
  \vev{\cdots}_S \equiv \int [d\phi] e^{S [\phi]} \cdots
\end{equation}
is the standard correlation function with the Boltzmann weight $e^{S
  [\phi]}$ given by the Wilson action. Let us
explain the role of the exponentiated second order differential by
considering the example of $n=2$.  We obtain
\begin{align}
  \vvev{\phi (p) \phi (q)}
  &= \frac{1}{K(p) K(q)} \left( \vev{\phi (p)
      \phi (q)} - \frac{K(p) \left(1 - K(p)\right)}{p^2} \delta
    (p+q)\right)\notag\\
  &= \frac{1}{K(p) K(q)} \left( \frac{K(p)}{p^2} -  \frac{K(p) \left(1
    - K(p)\right)}{p^2} \right) \delta (p+q) \notag\\
  &= \frac{1}{p^2} \delta   (p+q)\,.
\end{align}
The exponentiated differential modifies the high momentum behavior a
little so that the correct propagator is recovered from the Wilson
action.  This simple modification works not only for the free theory
but also for the interacting theory.\cite{Sonoda:2015bla}

A composite operator $\Op (p)$ with scale dimension $-y$ satisfies the
ERG differential equation
\begin{equation}
  \left( p \cdot \partial_p + y - \mathcal{D} \right) \Op (p) = 0\,,
  \label{sec2-ERG-Op}
\end{equation}
where
\begin{equation}
  \mathcal{D} \equiv \int_p \left[ \left( p \cdot \partial_p + \frac{D+2}{2}
    \right) \phi (p) \cdot \frac{\delta}{\delta \phi (p)} + f (p)
    \frac{1}{2} \frac{\delta^2}{\delta \phi (p) \delta \phi
    (-p)} \right]\,.
\end{equation}
$f(p)$ is the derivative of the high-momentum propagator $h(p)$:
\begin{subequations}
\begin{align}
  f (p) &\equiv \left( p \cdot \partial_p + 2 \right) h (p) =
          \frac{\Delta (p)}{p^2}\,,\label{sec2-f}\\
  h (p) &\equiv \frac{1 - K(p)}{p^2}\,.\label{sec2-h}
\end{align}
\end{subequations}
Please note that $-y$ is the scale dimension of the operator in
momentum space.  The scale dimension of the corresponding operator in
coordinate space is $D-y$, often denoted as $x$ (not to be confused
with space coordinate).  Eq.~(\ref{sec2-ERG-Op}) implies that the
correlation function of the composite operator
\begin{align}
&  \vvev{\Op (p) \phi (p_1) \cdots \phi (p_n)}\nt\\
  & \equiv \prod_{i=1}^n \frac{1}{K(p_i)}
    \vev{ \Op (p) \exp \left( -\int_q \frac{K(q)\left(1-K(q)\right)}{q^2}
    \frac{1}{2}     \frac{\delta^2}{\delta \phi (q)\delta \phi (-q)}\right)
    \phi (p_1) \cdots \phi (p_n)}_S \label{sec2-Op-corr}
\end{align}
satisfies the scaling relation
\begin{equation}
  \vvev{\Op (p e^t) \phi (p_1 e^t) \cdots \phi (p_n e^t)}
  = e^{ - \left( y + n \frac{D+2}{2} \right) t} \vvev{\Op (p) \phi
    (p_1) \cdots \phi (p_n)}\,.
\end{equation}
Note the absence of the factor $1/K(p)$ for the composite operator in
Eq.~(\ref{sec2-Op-corr}).  A composite operator acts like an infinitesimal
variation of the action $S$, and Eq.~(\ref{sec2-Op-corr}) can be regarded as the
corresponding infinitesimal variation of Eq.~(\ref{sec2-def-corr}).

We are particularly interested in the scalar composite operators which
are even in $\phi$ and relevant, i.e., the scale dimensions $-y$ are
either zero or negative.  There are only three independent operators:
$\Op_2$ corresponding to $\phi^2$, $\Op_4$ corresponding to $\phi^4$,
and $\N$ corresponding to $\partial_\mu \phi \partial_\mu \phi$.  (We
do not count $\partial^2 \phi^2$ separately.)  The purpose of this
paper is to understand the multiple products of the three composite
operators.  Let us end this section by constructing these composite
operators by solving the respective ERG differential equations.

\subsection{$\Op_2$ with $y=2$}

We introduce $\Op_2$ by the ERG equation
\begin{equation}
  \left( p \cdot \partial_p + 2 - \mathcal{D}\right) \Op_2 (p) = 0\,.
\end{equation}
This gives
\begin{equation}
  \Op_2 (p)
= \frac{1}{2} \int_{p_1, p_2} \phi (p_1) \phi (p_2) \, \delta
    \left( p_1 + p_2 - p \right) + v_2 \delta (p)\,,
\end{equation}
where
\begin{equation}
  v_2 \equiv - \frac{1}{4} \int_q f(q)\,.\label{sec2-v2}
\end{equation}
($f$ is defined by Eq.~(\ref{sec2-f}).)  We may graphically express
$\Op_2$ as
\begin{equation}
  \Op_2 (p) = \raisebox{-0.4cm}{\includegraphics{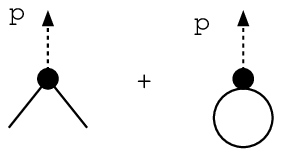}}\,,
\end{equation}
where each open solid line corresponds to a factor of $\phi$, and the
loop denotes $v_2$, corresponding to an integral over the
high-momentum propagator.\footnote{The naive expression for $v_2$ is
  $\frac{1}{2} \int_p h (p)$ which is quadratically divergent, but the
  ERG gives a perfectly finite expression (\ref{sec2-v2}).  A
  hand-waving derivation of this result is integration by parts,
  ignoring the divergent surface integral:
  $0 = \int_p (p \cdot \partial_p + 4) h(p) = \int_p f(p) + 2 \int_p h
  (p)$ gives
  $v_2 = \frac{1}{2} \int_p h(p) = - \frac{1}{4} \int_p f(p)$.}
We give graphs merely to aid the intuitive understanding of the
results, and they should not be taken too seriously.

\subsection{$\Op_4$ with $y=0$}

We introduce $\Op_4$ by
\begin{equation}
  \left( p \cdot \partial_p - \mathcal{D} \right) \Op_4 (p) = 0\,.
\end{equation}
This gives
\begin{align}
  \Op_4 (p)
  &= \frac{1}{4!} \int_{p_1, \cdots. p_4} \prod_{i=1}^4 \phi (p_i)\,
    \delta \left(\sum_{i=1}^4 p_i - p \right)\nt\\
  &\quad + v_2 \frac{1}{2} \int_{p_1, p_2} \phi (p_1) \phi (p_2)\,
    \delta (p_1+p_2-p) + \frac{1}{2} v_2^2 \delta (p)\\
  &= \raisebox{-0.6cm}{\includegraphics{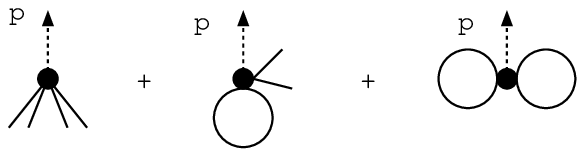}}\,.
\end{align}
In fact
\begin{equation}
  p^2 \Op_2 (p) = p^2 \frac{1}{2} \int_{p_1, p_2} \phi (p_1) \phi
  (p_2)\, \delta (p_1+p_2-p)
\end{equation}
satisfies the same ERG differential equation as $\Op_4 (p)$.

\subsection{$\N$ with $y=0$}

The number operator $\N$ is an equation-of-motion operator, defined by
\begin{equation}
  \N (p) \equiv - e^{-S} \int_q K(q) \frac{\delta}{\delta \phi (q)}
  \left( \phi (q+p) e^S \right)\,.\label{sec2-def-N}
\end{equation}
This is computed as
\begin{equation}
  \N (p)
  = \frac{1}{2} \int_{p_1, p_2} \phi (p_1) \phi (p_2) \, \left( p_1^2
    + p_2^2 \right)\, \delta (p_1+p_2-p) - \int_q K(q)\, \delta (p)\,.
\end{equation}
We will give a general discussion of the number operator and its
multiple products in Sec.~\ref{section-number}.

\section{Multiple products of composite operators\label{section-multiple}}

Multiple products of composite operators have been considered in \cite{Pagani:2017tdr}.
Here we review what is necessary for this paper.  For more details we
refer the reader to \cite{Pagani:2017tdr}.  See \cite{Pagani:2020ejb}
for a recent application of the formalism.  

Consider the product of a composite operator $\Op_i (p)$ of scale
dimension $- y_i$ and another $\Op_j (p)$ of scale dimension $- y_j$.\footnote{
We use an abstract notation in this section.  $\Op_2, \Op_4$, and $\N$ introduced in
the previous section are examples of $\Op_i$ discussed here.  We will
return to the concrete discussion of $\Op_2, \Op_4, \N$ in the next
section.}  Their product is given in the form
\begin{equation}
  \op{\Op_i (p) \Op_j (q)} = \Op_i (p) \Op_j (q) + \PP_{ij} (p, q)\,.
\end{equation}
The counterterm $\PP_{ij}$ is necessary for the product to be a composite
operator that satisfies
the ERG equation
\begin{equation}
  \left( p \cdot \partial_p + q \cdot \partial_q + y_i + y_j -
    \mathcal{D} \right)  \op{\Op_i (p) \Op_j (q)}
  = 0\,.\label{sec3-ERG-product}
\end{equation}
This is equivalent to the scaling law
\begin{equation}
  \vvev{\op{\Op_i (p e^t) \Op_j (q e^t)} \phi (p_1 e^t) \cdots \phi (p_n
    e^t)}
  = e^{- (y_i+y_j) t - n \frac{D+2}{2} t} \vvev{\op{\Op_i (p) \Op_j (q)}
    \phi (p_1) \cdots \phi (p_n)}\,.
\end{equation}
The ERG equation for the counterterm is obtained from
(\ref{sec3-ERG-product}) as
\begin{equation}
   \left( p \cdot \partial_p + q \cdot \partial_q + y_i + y_j -
    \mathcal{D} \right) \PP_{ij} (p, q) = \int_r f (r) \frac{\delta \Op_i
    (p)}{\delta \phi (r)} \frac{\delta \Op_j (q)}{\delta \phi (-r)}\,.
\end{equation}

The counterterm is induced by the short-distance singularity of the
product of two composite operators.  The singularity may induce
a mixing of the product with a local composite
operator $\Op_k$ with scale dimension $-y_k$ if
\begin{equation}
  -y_k + 2 d = - y_i - y_j \Longleftrightarrow y_k = y_i+y_j+ 2 d
\end{equation}
where $d = 0, 1, 2, \cdots$.   The ERG equation is then modified to
\begin{equation}
  \left( p \cdot \partial_p + q \cdot \partial_q + y_i + y_j -
    \mathcal{D}\right)
  \op{\Op_i (p) \Op_j (q)} = \gamma_{ij,k} (p,q) \Op_k
  (p+q)\,,\label{sec2-two-product}
\end{equation}
where $\gamma_{ij,k} (p,q)$ is a degree $2 d$ polynomial of
$p, q$.  Unless there is such $y_k$, there is no mixing.  We will give
examples in the next section.  Since we can rewrite
(\ref{sec2-two-product}) as
\begin{equation}
   \left( p \cdot \partial_p + q \cdot \partial_q + y_i + y_j -
    \mathcal{D}\right) \left(
  \op{\Op_i (p) \Op_j (q)} - \ln p \cdot \gamma_{ij,k} (p,q) \Op_k
  (p+q) \right) = 0\,,
\end{equation}
we find that the mixing causes the short-distance singularity
\begin{equation}
  \op{\Op_i (p) \Op_j (q)} \overset{p, q \to \infty}{\longrightarrow} \ln
  p \cdot \gamma_{ij,k} (p,q) \Op_k  (p+q) \,,\label{sec3-two-product-asymp}
\end{equation}
where $p+q$ is fixed.  Note that the solution of
(\ref{sec2-two-product}) is ambiguous: $\op{\Op_i (p) \Op_j (q)}$ can be
redefined with an addition of
\begin{equation}
  \sigma_{ij,k} (p,q)\,\Op_k (p+q)\label{sec3-two-product-redefinition}
\end{equation}
where $\sigma_{ij,k} (p,q)$ is a polynomial of degree $2d$.  But this
redefinition does not affect the logarithmic singularity
(\ref{sec3-two-product-asymp}).

We can consider higher order products.  The product of three composite
operators is given as
\begin{align}
  \op{\Op_i (p) \Op_j (q) \Op_k (r)}
  &= \Op_i (p) \Op_j (q) \Op_k (r) + \PP_{ij} (p,q) \Op_k (r)\nt\\
  &\quad + \PP_{jk} (q,r) \Op_i (p) + \PP_{ki} (r,p) \Op_j (q)\nt\\
  &\quad + \PP_{ijk} (p,q,r)\,,
\end{align}
where $\PP_{ijk}$ takes care of the short-distance
singularity that occurs when all the three operators come close
together in space.  Allowing for the possibility of mixing, we obtain
the ERG equation for the product as
\begin{align}
&  \left( p \cdot \partial_p + q \cdot \partial_q + r \cdot \partial_r
  + y_i + y_j + y_k - \mathcal{D} \right) \op{\Op_i (p) \Op_j (q) \Op_k (r)}
                \nt\\
  &= \gamma_{ij, l} (p,q) \op{\Op_l (p+q) \Op_k
    (r)} + \gamma_{jk, l} (q,r) \op{\Op_l (q+r)  \Op_i (p)}\nt\\
  &\quad + \gamma_{ki, l} (r,p) \op{\Op_l (r+p) \Op_j (q)} + \gamma_{ijk, l} (p,q,r) \Op_l
    (p+q+r)\,,\label{sec3-ERG-three-product}
\end{align}
where the mixing coefficient $\gamma_{ijk, l} (p,q,r)$ can be nonvanishing only if
\begin{equation}
  - y_i - y_j - y_k + y_l = 2 d' = 0, 2, 4, \cdots\,.
\end{equation}
Then it is a polynomial of degree $2d'$.  When all the three operators
have large momenta, we find
\begin{align}
  \op{\Op_i (p) \Op_j (q) \Op_k (r)}&\nt\\
  \overset{p, q, r \to  \infty}{\longrightarrow}
  &\frac{1}{2} \left( \ln p \right)^2 \cdot \Big\lbrace
   \gamma_{ij,l} (p,q)  \gamma_{lk, m} (p+q,r)+  \gamma_{jk,l} (q,r)
    \gamma_{li,m} (q+r,p) \nt\\
  &\qquad\quad + \gamma_{ki,l}
    (r,p)\gamma_{lj,m} (r+p,q) \Big\rbrace \, \Op_m (p+q+r)\nt\\
  &\quad + \ln p \cdot \left(\gamma_{ijk,l} (p,q,r) + \cdots \right)\,
    \Op_l (p+q+r)\,,\label{sec3-three-product-asymp} 
\end{align}
where $p+q+r$ is fixed.  Note that $\gamma_{ijk,l}$ is not the only
source of the term proportional to $\ln p$.  For example, the
short-distance expansion of $\op{\Op_l (p+q) \Op_k (r)}$ contains a
term $\tau_{lk,m} (p+q,r) \Op_m (p+q+r)$ free of $\ln p$, which
contributes a log term
\begin{equation}
  \ln p \cdot \gamma_{ij,m} (p,q) \tau_{mk,l} (p+q,r)
\end{equation}
as part of the dots in
(\ref{sec3-three-product-asymp}). ($\tau_{mk,l}$ is a polynomial of
degree $-y_m-y_k+y_l$.)  It is important to note that the mixing
coefficient $\gamma_{ijk,l}$ depends on the definition of the products
of two composite operators.  If we redefine the products of two
composite operators as (\ref{sec3-two-product-redefinition}),
$\gamma_{ijk,l} (p,q,r)$ changes accordingly.  We will derive the
changes shortly.

The ERG equation for $\PP_{ijk}$, corresponding to
(\ref{sec3-ERG-three-product}), is given by
\begin{align}
&  \left( p \cdot \partial_p + q \cdot \partial_q + r \cdot \partial_r
  + y_i + y_j + y_k - \mathcal{D} \right) \PP_{ijk} (p,q,r) \nt\\
  &= \int_s f(s) \left( \frac{\delta \PP_{ij} (p,q)}{\delta \phi
    (s)} \frac{\delta \Op_k (r)}{\delta \phi (-s)} +\frac{\delta
    \PP_{jk} (q,r)}{\delta \phi (s)} \frac{\delta \Op_i
    (p)}{\delta \phi (-s)} + \frac{\delta \PP_{ki} (r,p)}{\delta
    \phi (s)} \frac{\delta \Op_j (q)}{\delta \phi (-s)} \right)\nt\\
&\quad + \gamma_{ij, l} (p,q) \PP_{lk} (p+q,r)
 + \gamma_{jk, l} (q,r) \PP_{li} (q+r, p)\nt\\
  &\quad + \gamma_{ki, l} (r,p) \PP_{lj}
    (r+p, q) + \gamma_{ijk, l} (p,q,r) \Op_l  (p+q+r)\,.
\end{align}
We will calculate products of three composite operators in Appendix
\ref{appendix:three}.

\subsection*{Changes of $\gamma_{ijk,l}$ induced by
  (\ref{sec3-two-product-redefinition})}

Let us redefine the products of two composite operators as
\begin{equation}
  \op{\Op_i (p) \Op_j (q)}'
  \equiv \Op_i (p) \Op_j (q) + \PP'_{ij} (p,q)\,,
\end{equation}
where
\begin{equation}
  \PP'_{ij} (p,q) = \PP_{ij} (p,q) + \sigma_{ij,k} (p,q) \Op_k
  (p+q)\,,
\end{equation}
as given by (\ref{sec3-two-product-redefinition}).  We
must then define the products of three composite operators as
\begin{align}
  \op{\Op_i (p) \Op_j (q) \Op_k (r)}'
  &\equiv \Op_i (p) \Op_j (q) \Op_k (r) + \PP'_{ij} (p,q) \Op_k
    (r)\nt\\
  &\quad + \PP'_{jk} (q,r) \Op_i (p) + \PP'_{ki} (r,p) \Op_j (q)\nt\\
  &\quad + \PP'_{ijk} (p,q,r)\,,
\end{align}
where $\PP'_{ijk}$ must be defined so that the product is a composite
operator of scale dimension $-y_i-y_j-y_k$.  To find $\PP'_{ijk}$, we
write the original product of three as
\begin{align}
  \op{\Op_i (p) \Op_j (q) \Op_k (r)}
  &= \Op_i (p) \Op_j (q) \Op_k (r) + \PP'_{ij} (p,q) \Op_k (r)\nt\\
  &\quad + \PP'_{jk} (q,r) \Op_i (p) + \PP'_{ki} (r,p) \Op_j (q)\nt\\
  &\quad + \PP_{ijk} (p,q,r) - \sigma_{ij,m} (p,q) \Op_m (p+q) \Op_k
    (r)\nt\\
  &\quad - \sigma_{jk,m} (q,r) \Op_m (q+r) \Op_i (p) - \sigma_{ki,m}
    (r,p) \Op_m (r+p) \Op_j (q)\,.
\end{align}
The last two lines cannot be the desired $\PP'_{ijk} (p,q,r)$ since it
is nonlocal, involving products of independent pieces.  To make it
local, we define the new product as
\begin{align}
  \op{\Op_i (p) \Op_j (q) \Op_k (r)}'
  &\equiv \op{\Op_i (p)  \Op_j (q) \Op_k (r)} + \sigma_{ij,m} (p,q)
    \op{\Op_m (p+q) \Op_k (r)}\nt\\
  &\quad + \sigma_{jk,m} (q,r) \op{\Op_m (q+r) \Op_i (p)} +
    \sigma_{ki,m} (r,p) \op{\Op_m (r+p) \Op_j (q)}\nt\\
  &= \Op_i (p)  \Op_j (q) \Op_k (r)  + \PP'_{ij} (p,q) \Op_k (r)\nt\\
  &\quad + \PP'_{jk} (q,r) \Op_i (p) + \PP'_{ki} (r,p) \Op_j (q)
 + \PP'_{ijk} (p,q,r)\,,
\end{align}
where
\begin{align}
  \PP'_{ijk} (p,q,r)
  &\equiv \PP_{ijk} (p,q,r) + \sigma_{ij,m} (p,q) \PP_{mk} (p+q,
  r)\nt\\
  &\quad + \sigma_{jk,m} (q,r) \PP_{mi} (q+r,p) + \sigma_{ki,m} (r,p)
  \PP_{mj} (r+p, q)
\end{align}
is local as desired.  The new product, being the sum of composite operators,
is obviously a composite operator of scale dimension $-y_i-y_j-y_k$.

Now, the new product satisfies the ERG equation
\begin{align}
  & \left( p \cdot \partial_p + q \cdot \partial_q + r \cdot
    \partial_r + y_i+y_j+y_k - \mathcal{D} \right) \op{\Op_i (p) \Op_j
    (q) \Op_k (r)}'\nt\\
  &= \gamma_{ij,l} (p,q) \op{\Op_l (p+q) \Op_k}' + \gamma_{jk,l} (q,r)
    \op{\Op_l (q+r) \Op_i (p)}' \nt\\
  &\quad + \gamma_{ki,l} (r,p) \op{\Op_l (r+p)
    \Op_j (q)}' + \gamma'_{ijk,l} (p,q,r) \Op_l (p+q+r)\,,
\end{align}
where the new mixing coefficient is given by
\begin{align}
  \gamma'_{ijk,l} (p,q,r)
  &= \gamma_{ijk,l} (p,q,r)\nt\\
  &\quad + \sigma_{ij,m} (p,q) \gamma_{mk,l} (p+q,r) - \gamma_{ij,m}
    (p,q) \sigma_{mk,l} (p+q,r)\nt\\
  &\quad +  \sigma_{jk,m} (q,r) \gamma_{mi,l} (q+r,p) - \gamma_{jk,m}
    (q,r) \sigma_{mi,l} (q+r,p)\nt\\
  &\quad +  \sigma_{ki,m} (r,p) \gamma_{mj,l} (r+p,q) - \gamma_{ki,m}
    (r,p) \sigma_{mj,l} (r+p,q)\,.\label{sec3-new-mixing}
\end{align}
If there is an operator with scale dimension
\begin{equation}
  - y_l = -y_i-y_j-y_k - 2 d'\quad (d' = 0, 1, 2, \cdots)\,,
\end{equation}
we can redefine the three-operator product further by adding
\begin{equation}
  \sigma_{ijk,l} (p,q,r)\, \Op_l (p+q+r)\,,
\end{equation}
where $\sigma_{ijk,l}$ is a polynomial of degree $2d'$.  But this
redefinition does not affect $\gamma'_{ijk,l}$.

\section{Products of two composite operators\label{section-products-two}}

We consider the products of two composite operators, either $\Op_2$ or
$\Op_4$ defined in Sec.~\ref{section-operators}, in this section.  The
first two examples have been obtained in \cite{Pagani:2017tdr}.  We
include them here for completeness and the convenience of the reader.
Through these examples we wish to verify the claim that the
short-distance singularities are determined by the coefficients of
mixing, or vice versa.

\subsection{$\op{\Op_2 \Op_2}$}

We compute
\begin{equation}
  \op{\Op_2 (p) \Op_2 (q)} = \Op_2 (p) \Op_2 (q) + \PP_{22} (p,q)\,,
\end{equation}
where $\PP_{22}$ satisfies
\begin{equation}
  \left( p \cdot \partial_p + q \cdot \partial_q + 4 - \mathcal{D}
  \right) \PP_{22} (p,q) = \int_r f(r) \frac{\delta \Op_2 (p)}{\delta
    \phi (r)} \frac{\delta \Op_2 (q)}{\delta \phi (-r)} + \gamma_{22,0} \, \delta (p+q)\,.
\end{equation}
Let
\begin{equation}
  \PP_{22} (p,q)
  = \frac{1}{2} \int_{p_1, p_2} \phi (p_1) \phi (p_2)\, \delta
  (p_1+p_2-p-q)\, c_{22,2} (p,q; p_1, p_2) + c_{22,0} (p) \delta (p+q)\,.
\end{equation}
$\op{\Op_2 (p) \Op_2 (q)}$ being a composite operator of scale
dimension $-4$, it is ambiguous by a constant multiple of $\delta
(p+q)$.  We remove this ambiguity by adopting a convention
\begin{equation}
  c_{22,0} (0) = 0\,.\label{sec2-c220-zero}
\end{equation}

$c_{22,2}$ satisfies the ERG equation
\begin{equation}
  \left( p \cdot \partial_p + q \cdot \partial_q + \sum_{i=1}^2 p_i
    \cdot \partial_{p_i}  + 2 \right) c_{22,2} (p,q; p_1, p_2) = f(p_1-p) + f(p_2-p)\,.
\end{equation}
The solution is given by
\begin{equation}
  c_{22,2} (p,q; p_1, p_2) = h (p_1 - p) + h (p_2 - p) = \includegraphics{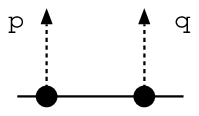}\,.
\end{equation}
where the solid line connecting two vertices corresponds to a
high-momentum propagator.

$c_{22,0}$ satisfies the ERG equation
\begin{equation}
  p \cdot \partial_p c_{22,0} (p) = \frac{1}{2} \int_q f(q) h (q+p) +
  \gamma_{22,0}\,.\label{sec4-F-diffeq} 
\end{equation}
Now, $c_{22,0} (p)$ must be analytic at $p=0$ since the fluctuations of
momentum below the momentum cutoff $1$ have not been integrated.  For the analyticity
at $p=0$, we must choose
\begin{equation}
  \gamma_{22,0} = - \int_q f(q) h(q) = - \frac{1}{(4 \pi)^2}\,.
\end{equation}
(This is calculated in Appendix \ref{appendix:integrals}.)  Using the
convention (\ref{sec2-c220-zero}), we obtain
\begin{equation}
  c_{22,0} (p) = F (p) = \raisebox{-0.3cm}{\includegraphics{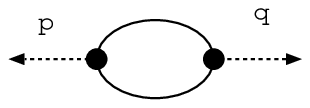}}\,,
\end{equation}
where $F(p)$ is defined by
\begin{equation}
  F(p) \equiv \frac{1}{2} \int_q h(q) \left( h (q+p) - h (q)
  \right)\,.\label{sec4-def-F}
\end{equation}
Note that $\gamma_{22,0}$ determines the asymptotic behavior of $F(p)$
for large $p$.  Since $F(p)$ satisfies
\begin{equation}
  p \cdot \partial_p F(p) \overset{p \to \infty}{\longrightarrow}
  \gamma_{22,0}\,,
\end{equation}
we obtain 
\begin{equation}
  F (p) \overset{p \to \infty}{\longrightarrow} \gamma_{22,0} \ln p
\end{equation}
up to an additive constant.  Hence, as $p, q$ become large with $p+q$
fixed, we find
\begin{equation}
  \op{\Op_2 (p) \Op_2 (q)} \overset{p, q \to \infty}{\longrightarrow}
  \gamma_{22,0} \ln p\, \delta (p+q)\,.
\end{equation}

\subsection{$\op{\Op_4 \Op_2}$}

$\op{\Op_4 (p) \Op_2 (q)}$ satisfies the ERG equation
\begin{equation}
  \left( p \cdot \partial_p + q \cdot \partial_q + 2 - \mathcal{D}
  \right) \op{\Op_4 (p) \Op_2 (q)} = \gamma_{42,2} \Op_2 (p+q) +
  \gamma_{42,0} (p) \delta (p+q)\,.
\end{equation}
Let
\begin{equation}
  \op{\Op_4 (p) \Op_2 (q)} = \Op_4 (p) \Op_2 (q) + \PP_{42} (p,q)\,.
\end{equation}
$\PP_{42}$ satisfies
\begin{equation}
  \left( p \cdot \partial_p + q \cdot \partial_q + 2 - \mathcal{D}
  \right) \PP_{42} (p,q) =
  \int_r f(r) \frac{\delta \Op_4 (p)}{\delta \phi (r)}
  \frac{\delta \Op_2 (q)}{\delta \phi (-r)} + \gamma_{42,2} \Op_2
  (p+q) + \gamma_{42,0} (p) \delta (p+q)\,.
\end{equation}
To solve this, we expand
\begin{align}
\PP_{42} (p,q) &= \frac{1}{4!} \int_{p_1, p_2, p_3, p_4} \prod_{i=1}^4
                 \phi (p_i) \, \delta \left( \sum_{i=1}^4 p_i - p -
                 q\right)\, c_{42,4} (p,q; p_1, \cdots, p_4)\nt\\
  &\quad + \frac{1}{2} \int_{p_1, p_2} \phi (p_1) \phi (p_2)\, \delta
    (p_1+p_2-p-q)\, c_{42,2} (p,q; p_1, p_2) + c_{42,0} (p) \delta (p+q)\,.
\end{align}

We first obtain
\begin{equation}
  \left(p \cdot \partial_p + q \cdot \partial_q + \sum_{i=1}^4 p_i
    \cdot \partial_{p_i} + 2 \right) c_{42,4} (p,q; p_1, 
  \cdots, p_4) =  \sum_{i=1}^4 f(p_i-q)\,,
\end{equation}
which is solved by
\begin{equation}
  c_{42,4} (p, q; p_1,\cdots,p_4)
  = \sum_{i=1}^4 h(p_i-q) = \raisebox{-0.4cm}{\includegraphics{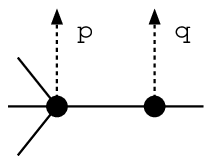}}\,.
\end{equation}

Let
\begin{align*}
  c_{42,2} (p, q; p_1, p_2)
  &= v_2 \left( h(p_1-q) + h(p_2-q) \right) + c_{42,2}^{\mathrm{1PI}}
    (p, q)\\
&= \raisebox{-0.4cm}{\includegraphics{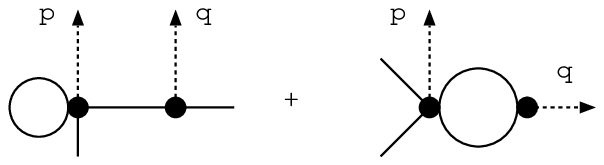}}\,,
\end{align*}
where $c_{42,2}^{\mathrm{\mathrm{1PI}}} (p,q)$ satisfies
\begin{equation}
  \left( p \cdot \partial_p + q \cdot \partial_q \right) c_{42,2}^{\mathrm{1PI}}
  (p, q) =  \int_r f(r) h (r+q) + \gamma_{42,2}\,.\label{sec4-diffeq-c422-1PI}
\end{equation}
For this to have a solution analytic at zero momenta, we must choose
\begin{equation}
  \gamma_{42,2} = - \int_r f(r) h(r) = \gamma_{22,0} = - \frac{1}{(4 \pi)^2}\,.
\end{equation}
The solution is ambiguous by a constant, i.e., we can change
$\op{\Op_4 (p) \Op_2 (q)}$ by a constant multiple of $\Op_2 (p+q)$.
Note Eq.~(\ref{sec4-diffeq-c422-1PI}) is the same as
Eq.~(\ref{sec4-F-diffeq}).  Hence, by adopting a convention
\begin{equation}
  c_{42,2}^{\mathrm{1PI}} (0,0) = 0\,,
\end{equation}
we obtain
\begin{equation}
  c_{42,2}^{\mathrm{1PI}} (p, q) = F(q)\,,
\end{equation}
where $F$ is defined by (\ref{sec4-def-F}).

Finally, we consider $c_{42,0} (p)$ that satisfies
\begin{equation}
   \left( p \cdot \partial_p - 2 \right) c_{42,0} (p) = \frac{1}{2} \int_r
  f(r) \cdot F(p) + v_2 \left( \int_r f(r) h(r+p) + \gamma_{42,2}
  \right) + \gamma_{42,0} (p)\,,
\end{equation}
where $\gamma_{42,0} (p)$ is a constant multiple of $p^2$.  For
$c_{42,0} (p)$ to be analytic at $p=0$, the rhs should have no term
proportional to $p^2$.  This determines
\begin{equation}
  \gamma_{42,0} (p) = 0\,.
\end{equation}
The solution is ambiguous by a constant multiple of $p^2$.
A particular solution is given by
\begin{equation}
  c_{42,0} (p) = v_2 F(p) = \raisebox{-0.4cm}{\includegraphics{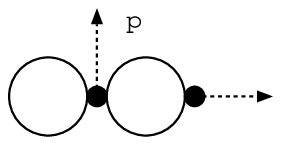}}\,.
\end{equation}

For large $p$ with $p+q$ fixed, we find the asymptotic behavior
\begin{equation}
  \op{\Op_4 (p) \Op_2 (q)} \overset{p, q \to \infty}{\longrightarrow}
  \gamma_{42,2} \ln p \cdot \Op_2 (p+q)\,.
\end{equation}

\subsection{$\op{\Op_4 \Op_4}$}

This example is not given in \cite{Pagani:2017tdr}, and we would like
to give more details than for the preceding examples.  The ERG
equation for the product is
\begin{align}
  & \left( p \cdot \partial_p + q \cdot \partial_q - \mathcal{D}
    \right) \op{\Op_4 (p) \Op_4 (q)}\nt\\
  &= \gamma_{44,4} \Op_4 (p+q) + \gamma_{44,\N} \N (p+q) +
    \gamma_{44,2} (p,q) \Op_2 (p+q) + \gamma_{44,0} (p) \delta (p+q)\,,
\end{align}
where $\gamma_{44,2} (p,q)$ is a degree $2$ polynomial, and
$\gamma_{44,0} (p)$ is proportional to $(p^2)^2$.  The corresponding
ERG equation for the counterterm is
\begin{align}
  & \left( p \cdot \partial_p + q \cdot \partial_q - \mathcal{D}
    \right) \PP_{44} (p,q)\nt\\
&= \int_r f(r) \frac{\delta \Op_4 (p)}{\delta \phi (r)} \frac{\delta
    \Op_4 (q)}{\delta \phi (-r)} + \gamma_{44,4} \Op_4 (p+q) +
                                  \gamma_{44,\N} \N (p+q) \nt\\
  &\quad +  \gamma_{44,2} (p,q) \Op_2 (p+q) + \gamma_{44,0} (p) \delta (p+q)\,.
\end{align}
The solution is ambiguous by a linear combination of
$\Op_4 (p+q), \N (p+q), (p\cdot q)\Op_2 (p+q)$, and $p^4 \delta (p+q)$
with constant coefficients, and we will introduce particular
conventions to remove the ambiguities.  We expand
\begin{align}
\PP_{44} (p,q) 
  &= \sum_{n=1}^3 \frac{1}{(2n)!} \int_{p_1, \cdots, p_{2n}}
    \prod_{i=1}^{2n} \phi (p_i)\, \delta \left( \sum_{i=1}^{2n} p_i -
    p - q\right)\, c_{44, 2n} (p,q; p_1, \cdots, p_{2n})\nt\\
  &\quad + \delta (p+q) \, c_{44,0} (p)\,.
\end{align}

The calculations of $c_{44,6}$ and $c_{44,4}$ are similar to what we
have shown for the previous two examples.  We simply state the
results:
\begin{align}
  c_{44,6} (p,q; p_1,\cdots, p_6) &=
                          \raisebox{-0.4cm}{\includegraphics{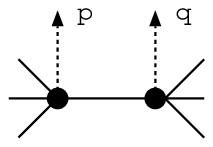}}\nt\\
                        &= h (p_1+p_2+p_3-p) + \cdots\,.\\
  c_{44,4} (p,q; p_1,\cdots,p_4) &=
                         \raisebox{-0.4cm}{\includegraphics{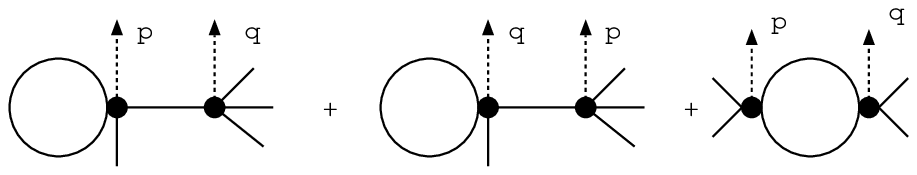}}\nt\\
                        &= v_2 \sum_{i=1}^4 \left( h(p_i-p) + h (p_i -
                          q) \right)  + c_{44,4}^{\mathrm{1PI}} (p,q; p_1, \cdots, p_4)\,,
\end{align}
where
\begin{equation}
   c_{44,4}^{\mathrm{1PI}} (p, q; p_1, \cdots, p_4) = F (p_1+p_2-p) +
   F (p_1 +p_2 - q) + \textrm{(t, u)-channels}\,.
\end{equation}
We have chosen
\begin{equation}
  \gamma_{44,4} = - 6 \int_q f(q) h(q) = 6 \gamma_{42,2} = -
  \frac{6}{(4 \pi)^2}
\end{equation}
for the analyticity of $c_{44,4}^{\mathrm{1PI}}$ at zero momenta, and we have chosen a
particular convention
\begin{equation}
  c_{44,4}^{\mathrm{1PI}} (0,0; 0,0,0,0) = 0
\end{equation}
to remove the ambiguity of $c_{44,4}^{\mathrm{1PI}}$ by a constant.  For
large $p, q$ with $p+q$ fixed, we find
\begin{equation}
  c_{44,4} (p, q; p_1, p_2, p_3, p_4) \overset{p, q \to \infty}{\longrightarrow}
  \gamma_{44,4} \ln p\,.
\end{equation}

We can write $c_{44,2}$ as
\begin{align}
  c_{44,2} (p,q; p_1, p_2) &= \raisebox{-0.4cm}{\includegraphics{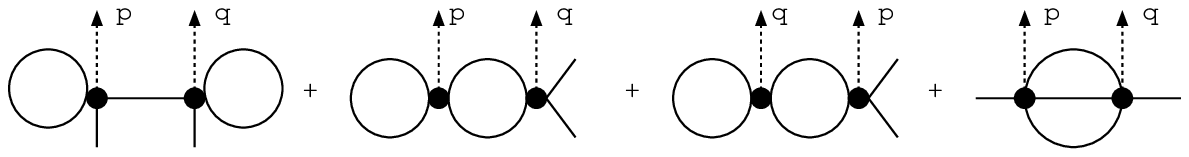}}\nt\\
                      &= v_2^2 \left( h (p_1 - p) + h (p_2 - p)
                        \right) + v_2 \left( F (p)  + F (q) \right) +
                        c_{44,2}^{\mathrm{1PI}} (p, q; p_1, p_2)\,,
\end{align}
where $c_{44,2}^{\mathrm{1PI}}$ satisfies the ERG equation
\begin{align}
&  \left( p \cdot \partial_p + q \cdot \partial_q + \sum_{i=1}^2 p_i
                \cdot \partial_{p_i} - 2 \right)
                c_{44,2}^{\mathrm{1PI}} (p,q; p_1, p_2)\nt\\
  &= \int_r f(r) \left( F (r+p_1-p) + F (r+p_2-p) \right) +
    \frac{2}{3} \gamma_{44,4} v_2 + \gamma_{44,\N} (p_1^2 + p_2^2) +
    \gamma_{44,2} (p,q)\,.
\end{align}
$\gamma_{44,\N}$ and the quadratic polynomial $\gamma_{44,2}
(p,q)$ are to be determined so that the rhs has no term of quadratic
order at zero momenta.  This is required by the analyticity of
$c_{44,2}^{\mathrm{1PI}} (p,q; p_1, p_2)$.

To solve this we define $G(p)$ by
\begin{equation}
  \left( p \cdot \partial_p - 2 \right) G (p)
  = \int_q f(q) F(q+p) + \frac{1}{3} \gamma_{44,4} v_2 + \eta \,
  p^2\,,\label{sec4-G-diffeq}
\end{equation}
where
\begin{equation}
  \eta \equiv - \frac{\partial}{\partial p^2} \int_q f(q) F(q+p)
  \Big|_{p=0} = \frac{1}{(4 \pi)^2} \frac{1}{6}
\end{equation}
removes the $p^2$ term from the rhs.  See Appendix
\ref{appendix:integrals} for the calculation of $\eta$.  We also
remove the ambiguity of $G (p)$ by 
the convention
\begin{equation}
  \frac{d}{dp^2} G(p)\Big|_{p=0} = 0\,.
\end{equation}
Since
\begin{align}
  & \left( p \cdot \partial_p + q \cdot \partial_q + \sum_{i=1}^2 p_i
    \cdot \partial_{p_i} - 2 \right) \left( G (p_1-p) +
    G(p_1 - q) \right)\nt\\
  &= \int_r f(r) \left( F (r+p_1-p) + F (r+p_1-q) \right) +
    \frac{2}{3} \gamma_{44,4} v_2 + \eta \lb (p_1-p)^2 + (p_2-p)^2
    \rb\nt\\
  &= \int_r f(r) \left( F (r+p_1-p) + F (r+p_1-q) \right) +
    \frac{2}{3} \gamma_{44,4} v_2 + \eta \left( p_1^2 + p_2^2 \right)
    - 2 \eta  (p \cdot q)\,,
\end{align}
we obtain
\begin{equation}
  c_{44,2}^{\mathrm{1PI}} (p, q; p_1, p_2) = G (p_1-p) + G (p_2 - p)\,,
\end{equation}
and
\begin{subequations}
  \begin{align}
    \gamma_{44,\N} &= \eta\,,\\
    \gamma_{44,2} (p,q) &= - 2 \eta \,p \cdot q\,.
  \end{align}
\end{subequations}
To summarize, we have obtained
\begin{align}
  c_{44,2} (p,q; p_1, p_2)
  &= v_2^2 \left( h (p_1-p) + h (p_2-p) \right) + v_2
    \left( F(p) + F (q)\right) \notag\\
  &\quad + G (p_1-p) + G(p_2-p)\,.
\end{align}

Before computing $c_{44,0} (p)$, let us find the asymptotic
behavior of $c_{44,2} (p, q; p_1, p_2)$.  The asymptotic behavior of $G(p)$ is obtained
in Appendix \ref{appendix:H} as (\ref{appendix-G-asymp}):
\begin{equation}
  G (p) \overset{p \to \infty}{\longrightarrow} \eta p^2 \ln p + 2 \gamma_{42,2} v_2 \ln p\,,
\end{equation}
where we have dropped a constant multiple of $p^2$ and a constant.
Hence, for large $p, q$ with fixed $p+q$, we obtain the asymptotic
behavior
\begin{equation}
  c_{44,2} (p, q; p_1, p_2)  \overset{p \to
    \infty}{\longrightarrow} \ln p \cdot \left[ \eta \, \left( - 2 p \cdot q + p_1^2 +
    p_2^2 \right) + \gamma_{44,4} v_2 \right]\,.
\end{equation}
  
Finally, we consider
\begin{align}
  c_{44,0} (p)
  &= \raisebox{-0.7cm}{\includegraphics{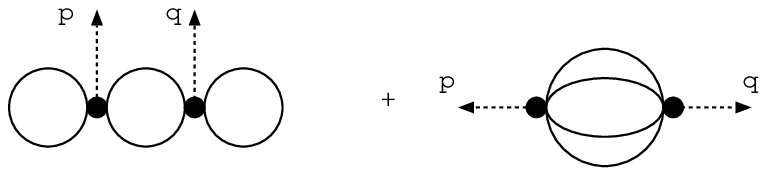}}\nt\\
  &= v_2^2 F (p) +  H (p)\,,
\end{align}
where $H(p)$ is defined by
\begin{equation}
  \left( p \cdot \partial_p - 4 \right) H (p)
  = \int_q f(q) G (q+p) + v_2^2 \frac{1}{3} \gamma_{44,4} - \eta \int_q K(q)
    + 2 \eta\, p^2 v_2 + \gamma_{44,0} (p)\,,\label{sec4-H-diffeq}
\end{equation}
and the convention
\begin{equation}
  \frac{d^2}{(d p^2)^2} H (p)\Big|_{p=0} = 0\,.
\end{equation}
$\gamma_{44,0} (p)$ is determined so that the rhs has no term of order
$p^4$ at $p=0$:
\begin{equation}
  \gamma_{44,0} (p) = \gamma_{44,0} \cdot p^4\,,
\end{equation}
where
\begin{equation}
  \gamma_{44,0} = - \frac{1}{2} \frac{d^2}{(d p^2)^2} \int_q f(q)
  G(q+p) \Big|_{p=0} = - \frac{1}{(4 \pi)^6} \frac{1}{144}\,.
\end{equation}
This is calculated in Appendix \ref{appendix:integrals}.
In Appendix \ref{appendix:H} we derive the asymptotic behavior of $H
(p)$ as
\begin{equation}
  H (p) \overset{p \to \infty}{\longrightarrow} \ln p \left[
  \gamma_{44,0} \, p^4 + 2 \eta\,  v_2 p^2  + 2
  \gamma_{42,2} v_2^2 \right]\,,
\end{equation}
where we have ignored constant multiples of $p^4, p^2, 1$.

To summarize, the asymptotic behavior of the operator product is given by
\begin{align}
  &  \op{\Op_4 (p) \Op_4 (q)} \nt\\
  &\overset{p, q \to \infty}{\longrightarrow}
\ln p \left(  \gamma_{44,4} \Op_4 (p+q) + \gamma_{44,\N} \N (p+q)  +
    \gamma_{44,2} (p,q)  \Op_2 (p+q) + \gamma_{44,0} \, p^4 \,\delta
  (p+q) \right)\,.
\end{align}

\section{Number Operator and its Multiple Products\label{section-number}}

The number operator defined by (\ref{sec2-def-N}) is an
equation-of-motion operator, and it has the correlation function
\begin{equation}
  \vvev{\N (p)\, \phi (p_1) \cdots \phi (p_n)}
  = \sum_{i=1}^n \vvev{\phi (p_1) \cdots \phi (p_i+p) \cdots \phi
    (p_n)}\,.\label{sec5-N-corr}
\end{equation}
See Appendix \ref{appendix:eom} for a quick review of the equation-of-motion composite
operators in the ERG formalism.

Given a composite operator $\Op$, we can create an equation-of-motion
operator by
\begin{equation}
\N (p) \star \Op \equiv - e^{-S} \int_q K(q) \frac{\delta}{\delta
    \phi (q)} \left( \op{\phi (q+p) \Op} e^S \right)\,,
\end{equation}
where
\begin{equation}
  \op{\phi (q+p) \Op} = \op{\Op \phi (q+p)} \equiv \phi (q+p) \Op + h (q+p)
  \frac{\delta \Op}{\delta \phi (-q-p)}
\end{equation}
is the product of $\phi$ and $\Op$ with the correlation function
\begin{equation}
  \vvev{\op{\Op \phi (q)} \phi (p_1) \cdots \phi (p_n)}
= \vvev{\Op\,\phi (q) \phi (p_1) \cdots \phi (p_n)}\,.
\end{equation}
$\N (p) \star \Op$ has the correlation function
\begin{equation}
  \vvev{\N (p) \star \Op\,\phi (p_1) \cdots \phi (p_n)} = \sum_{i=1}^n
  \vvev{\Op\, \phi (p_1) \cdots \phi (p_i+p)\cdots \phi (p_n)}\,,
\end{equation}
where $\Op$ is untouched by the number operator.

By definition the star product is commuting:
\begin{equation}
  \N (p) \star \N (q) \star \Op = \N (q) \star \N (p) \star \Op\,.
\end{equation}
This is easy to prove from the correlation functions:
\begin{align}
& \vvev{\N (p) \star \N (q) \star \Op\, \phi (p_1) \cdots \phi
                (p_n)}  = \sum_{i=1}^n \vvev{\N (q) \star \Op\, \cdots \phi (p_i+p)
    \cdots}\nt\\
  &= \sum_{i=1}^n \vvev{\Op\, \cdots \phi (p_i+p+q) \cdots} + \sum_{i
    \ne j} \vvev{\Op\, \cdots \phi (p_i+p) \cdots \phi 
    (p_j+q) \cdots}\,.
\end{align}
This is obviously symmetric under the interchange of $p$ and $q$.

Given
\[
  \Op = \op{\Op_2 (p_1) \cdots \Op_2 (p_k) \Op_4 (q_1) \cdots \Op_4
    (q_l)}\,,
\]
we define its product with a multiple number of $\N$'s as
\begin{align}
&  \op{\N (p_1) \cdots \N (p_k) \, \Op_2 (p_1) \cdots \Op_2 (p_k) \Op_4 (q_1) \cdots \Op_4
    (q_l)} \nt\\
&  \equiv \N (p_1) \star \cdots \star \N (p_k) \star  \op{\Op_2 (p_1)
                   \cdots \Op_2 (p_k) \Op_4 (q_1) \cdots \Op_4
                   (q_l)}\,.
                   \label{sec5-multiple-N}
\end{align}
According to this definition, we find that the product with $\N$ does
not produce any new short-distance singularity.  Suppose the product
$\op{\Op (p) \Op' (q)}$ satisfies the ERG equation
\begin{equation}
  \left( p \cdot \partial_p + q \cdot \partial_q + y_i + y_j -
    \mathcal{D} \right) \op{\Op_i (p) \Op_j (q)} = \gamma_{ij, k}
  (p,q) \Op_k (p+q)\,.
\end{equation}
The rhs is due to the short-distance singularity.  Taking the star product with
$\N (r)$, we obtain
\begin{equation}
  \left( p \cdot \partial_p + q \cdot \partial_q + r \cdot \partial_r + y_i + y_j -
    \mathcal{D} \right) \op{\N (r) \Op_i (p) \Op_j (q)} = \gamma_{ij, k}
  (p,q) \op{\N (r) \Op_k (p+q)}\,.
\end{equation}
$\N$ itself has no short-distance singularity with any composite
operator.  Thus, the only mixing coefficient involving $\N$ is of the type
\begin{equation}
  \gamma_{4 \cdots 4,\, \N}\,.
\end{equation}

Let us consider two examples.  The first example is
\begin{align}
  \op{\N (p) \Op_2 (q)}
  &= \N (p) \star \Op_2 (q)\nt\\
  &= - e^{-S} \int_r K(r) \frac{\delta}{\delta \phi (r)} \left[
    \left( \phi (r+p) \Op_2 (q) + h (r+p) \frac{\delta \Op_2
    (q)}{\delta \phi (-r-p)} \right) e^S \right]\nt\\
  &= \N (p) \Op_2 (q) + \PP_{\N 2} (p,q)\,,
\end{align}
where
\begin{align}
  \PP_{\N 2} (p,q)
  &= \int_r \frac{\delta \N (p)}{\delta \phi (r)} h(r) \frac{\delta
    \Op_2 (q)}{\delta \phi (-r)} \nt\\
  &\quad - 2 \Op_2 (p+q) + \left( 2 v_2 -
    \int_r K(r) h (r+p) \right)\, \delta (p+q)\,.
\end{align}
The second example is
\begin{align}
  \op{\N (p) \Op_4 (q)}
  &= \N (p) \star \Op_4 (q)\nt\\
  &= - e^{-S} \int_r K(r) \frac{\delta}{\delta \phi (r)} \left[ \left(
    \phi (r+p) \Op_4 (q) + h (r+p) \frac{\delta \Op_4 (q)}{\delta \phi
    (-r-p)} \right) e^S \right]\nt\\
  &= \N (p) \Op_4 (q) + \PP_{\N 4} (p,q)\,,
\end{align}
where
\begin{align}
  \PP_{\N 4} (p,q)
  &= \int_r \frac{\delta \N (p)}{\delta \phi (r)} h (r) \frac{\delta
    \Op_4 (q)}{\delta \phi (-r)}\nt\\
  &\quad - 4 \Op_4 (p+q) + \left( 4 v_2 - \int_r K(r) h(r+p) \right)
    \Op_2 (p+q)\nt\\
  &\quad - v_2 \frac{1}{2} \int_{p_1,p_2} \phi (p_1) \phi (p_2) \delta
    (p_1+p_2-p-q)\, \left( 1 - K(p_1-p) + 1 - K(p_2-p)\right)\nt\\
  &\quad - 2 v_2^2\, \delta (p+q)\,.
\end{align}

\section{Generating functional\label{section-W}}

So far we have shown how to construct the multiple products of $\Op_2$
and $\Op_4$ by solving the respective ERG differential equations.
Their products with the number operator $\N$ are given by star
products as discussed in the previous section.

To better understand the relation among the multiple products, we
couple momentum dependent sources to $\Op_2, \Op_4, \N$ and introduce
a generating functional $W [J_2, J_4, J_\N]$ by
\begin{align}
  W [J_2, J_4, J_\N]
  &\equiv \int_p \left( J_2 (-p) \Op_2 (p) + J_4 (-p) \Op_4 (p) + J_\N
    (-p) \N (p) \right)\nt\\
  &\quad + \sum_{k+l+m\ge 2} \frac{1}{k!\, l! \,m!} \int_{p_1,\cdots,p_k;
    q_1, \cdots, q_l; r_1, \cdots, r_m} J_2 (-p_1) \cdots J_2
    (-p_k)\nt\\
  &\qquad \times  J_4 (-q_1) \cdots J_4 (-q_l) J_\N (-r_1) \cdots J_\N (-r_m)\nt\\
  &\qquad \times \PP_{2\cdots 4 \cdots \N \cdots}
    (p_1,\cdots,p_k,q_1,\cdots, q_l, r_1 \cdots, r_m)\,.
\end{align}
Its exponential $e^{W [J_2, J_4, J_\N]}$ is the sum of composite operators:
\begin{align}
  e^{W [J_2,J_4,J_\N]}
 &= 1 + \sum_{k+l+m\ge 1} \frac{1}{k!\, l!\, m!}  \int_{p_1,\cdots,p_k;
    q_1, \cdots, q_l; r_1, \cdots, r_m} J_2 (-p_1) \cdots J_2
    (-p_k)\nt\\
  &\qquad\qquad \times  J_4 (-q_1) \cdots J_4 (-q_l) J_\N (-r_1) \cdots J_\N
    (-r_m)\nt\\
  &\qquad\qquad \times \op{\Op_2 (p_1) \cdots \Op_2 (p_k) \Op_4 (q_1) \cdots
    \Op_4 (q_l) \N (r_1) \cdots \N (r_m)}\,.
\end{align}
We call $W$ the generating functional because the multiple products
are obtained as its differentials:
\begin{align}
&  \op{\Op_2 (p_1) \cdots \Op_2 (p_k) \Op_4 (q_1) \cdots \Op_4 (q_l) \N
    (r_1) \cdots \N (r_m)}\nt\\
&= \frac{\delta^{k+l+m}}{\delta J_2 (-p_1)
    \cdots \delta J_2 (-p_k) \delta J_4 (-q_1) \cdots \delta J_4
    (-q_l) J_\N (-r_1) \cdots J_\N (-r_m)} e^{W [J_2, J_4,
                                  J_\N]}\Big|_{J_2=J_4=J_\N = 0}\,.
\end{align}

By giving the scale dimensions $-2, -4, -4$ to
$J_2 (-p), J_4 (-p), J_\N (-p)$, respectively, we can make
$e^{W[J_2,J_4,J_\N]}$ a composite operator of scale dimension $0$.  We
can then write the ERG differential equation compactly as
\begin{align}
&  \int_p \left[ \lb \left( - p \cdot \partial_p - 2 \right) J_2 (-p) - B_2
                (-p) \rb  \frac{\delta}{\delta J_2 (-p)}\nt\right.\\
&\quad + \lb \left( - p \cdot \partial_p - 4 \right) J_4 (-p) - B_4
    (-p) \rb \frac{\delta}{\delta J_4 (-p)}\label{sec6-W-ERG}\\
  &\left.\quad + \lb \left( - p \cdot \partial_p - 4 \right) J_\N (-p) - B_\N
    (-p) \rb \frac{\delta}{\delta J_\N (-p)} - \mathcal{D} \right] e^{W [J_2, J_4,
    J_\N]} = B_0 [J_2,J_4]\, e^{W[J_2, J_4, J_\N]}\,,\nt
\end{align}
where
\begin{align}
&  B_0 [J_2, J_4]
  \equiv \sum_{k=0}^\infty \frac{1}{k!} \gamma_{4 \cdots 4
    2 2, 0} \int_{p_1, \cdots, p_k} J_4 (-p_1) \cdots J_4
    (-p_k) \nt\\
  &\qquad\qquad\qquad \times \frac{1}{2} \int_{q_1, q_2} J_2 (-q_1) J_2 (-q_2)\, \delta
    \left( p_1 + \cdots + p_k + q_1 + q_2 \right)\nt\\
  &\quad + \sum_{k=1}^\infty \frac{1}{k!} \int_{p_1, \cdots, p_k}
    \gamma_{4 \cdots 4 2, 0} (p_1, \cdots, p_k, q) J_4
    (-p_1) \cdots J_4 (-p_k) J_2 (-q) \delta (p_1 + \cdots + p_k +
    q)\nt\\
  &\quad + \sum_{k=2}^\infty \frac{1}{k!} \int_{p_1, \cdots, p_k}
    \gamma_{4 \cdots 4, 0} (p_1, \cdots, p_k)\, J_4
    (-p_1) \cdots J_4 (-p_k)\, \delta (p_1 + \cdots + p_k)\,,\label{sec6-B0}\\
  B_2 (-p)
  &\equiv \sum_{k=1}^\infty \frac{1}{k!} \gamma_{4\cdots 4
    2, 2} \int_{p_1,\cdots,p_k,q} J_4 (-p_1) \cdots J_4 (-p_k) J_2
    (-q)\, \delta (p_1 + \cdots + p_k + q - p)\nt\\
  &\quad + \sum_{k=2}^\infty \frac{1}{k!} \int_{p_1, \cdots, p_k}
    \gamma_{4 \cdots 4, 2} (p_1, \cdots, p_k) J_4
    (-p_1) \cdots J_4 (-p_k) \,\delta (p_1 + \cdots p_k - p) \,,\label{sec6-B2}\\
  B_4 (-p)
  &\equiv \sum_{k=2}^\infty \frac{1}{k!} \gamma_{4 \cdots 4,\,
    4} \int_{p_1, \cdots, p_k} J_4 (-p_1) \cdots J_4 (-p_k)\, \delta
    (p_1 + \cdots + p_k - p)\,,\label{sec6-B4}\\
  B_\N (-p)
  &\equiv  \sum_{k=2}^\infty \frac{1}{k!} \gamma_{4 \cdots 4,\,\N} \int_{p_1,
    \cdots, p_k} J_4 (-p_1) \cdots J_4 (-p_k)\, \delta 
    (p_1 + \cdots + p_k - p)\,.\label{sec6-BN}
\end{align}
Our notation for the mixing coefficients may be clear.  For
example, the product of $k$ number of $\Op_4$ satisfies the ERG
equation
\begin{align}
&  \left( \sum_{i=1}^k p_i \cdot \partial_{p_i} - \mathcal{D} \right)
                \op{\Op_4 (p_1) \cdots \Op_4 (p_k)} \nt\\
  &= \gamma_{4 \cdots 4, \,4} \Op_4 (p_1+\cdots + p_k) + \gamma_{4
    \cdots 4, \,\N} \N (p_1+\cdots + p_k)\nt\\
  &\quad + \gamma_{4 \cdots 4, \,2} (p_1, \cdots, p_k) \Op_2 (p_1 +
    \cdots + p_k) + \gamma_{4 \cdots 4, \,0} (p_1, \cdots, p_k) \,
    \delta (p_1 + \cdots + p_k) + \cdots\,,
\end{align}
where the mixing of less than $k$ number of $\Op_4$'s is suppressed.
As we have discussed in Sec.~\ref{section-multiple}, the mixing
coefficients determine the short-distance singularity of the operator
product.  When all the momenta become large with the sum
$\sum_{1}^k p_i$ fixed, the short-distance singularity is given by
\begin{align}
&  \op{\Op_4 (p_1) \cdots \Op_4 (p_k)}\nt\\
  \overset{p_1, \cdots, p_k \to
  \infty}{\longrightarrow}
  &\frac{k!}{2^{k-1}} (\ln p)^{k-1}  \gamma_{44,4}^{k-2} \Bigg[
                               \gamma_{44,4} \Op_4  \left(\sum
    p_i\right) + \gamma_{44,\N} \N  \left(\sum p_i\right) +
\nt\\
  &\qquad + \frac{1}{k} \sum_{n=1}^k \gamma_{44,2} \left(\sum p_i - p_n,
    p_n\right) \Op_2 \left(\sum p_i\right) \nt\\
  &\qquad + \frac{1}{k} \sum_{n=1}^k \gamma_{44,0} \left(\sum p_i
    - p_n, p_n\right) \delta \left(\sum p_i\right) \Bigg]\nt\\
  & + \cdots\nt\\
      & +
  \ln p\, \Big[ \gamma_{4 \cdots 4, \,4} \Op_4 \left(\sum p_i\right) + \gamma_{4
    \cdots 4, \,\N} \N \left(\sum p_i\right)\nt\\
&\quad + \gamma_{4 \cdots 4, \,2} (p_1, \cdots, p_k) \Op_2 \left(\sum p_i\right)
 + \gamma_{4 \cdots 4, \,0} (p_1, \cdots, p_k) \,
                                         \delta \left(\sum p_i\right) + \cdots \Big]\,,
\end{align}
where $p$ is any of the momenta $p_i$.   The dotted parts,
proportional to the powers of $\ln p$ less than $k-1$, are
determined by the lower order mixing coefficients.

To convince ourselves that (\ref{sec6-W-ERG}) is correct, it may
suffice to check three terms.  The term first order in $J_2$ gives
\begin{equation}
  \int_p \left[ \left( - p \cdot \partial_p - 2 \right) J_2 (-p) - J_2
                (-p) \mathcal{D}\right]  \Op_2 (p)
= \int_p J_2 (-p) \left( p \cdot \partial_p + 2 - \mathcal{D}
    \right) \Op_2 (p) = 0\,,
\end{equation}
which is correct.  Similarly, the term first order in $J_4$ gives
\begin{equation}
 \int_p \left[ \left( - p \cdot \partial_p - 4 \right) J_4 (-p) -
    J_4 (-p) \mathcal{D}\right] \Op_4 (p)
= \int_p J_4 (-p) \left( p \cdot \partial_p - \mathcal{D} \right)
    \Op_4 (p) = 0\,,
\end{equation}
which is correct again.  The term proportional to $J_4 J_2$ gives
\begin{align}
&  \int_{p,q} \Bigg[ \left( - p \cdot \partial_p - 2 \right) J_2 (-p) \cdot J_4
                (-q) \,\op{\Op_2 (p) \Op_4 (q)} \nt\\
  &\qquad + J_2 (-p) \left(- q \cdot \partial_q - 4
                \right) J_4 (-q)\cdot \op{\Op_2 (p) \Op_4
                (q)} - \gamma_{42,2}\, J_2 (-p) J_4 (-q)\, \Op_2 (p+q) \Bigg]\nt\\
&= \int_{p,q} \gamma_{42,0} (q,p) J_2 (-p) J_4 (-q) \,\delta (p+q)\,.
\end{align}
This gives the correct equation
\begin{equation}
 \left( p \cdot \partial_p + q
  \cdot \partial_q + 2 - \mathcal{D} \right) \op{\Op_2 (p) \Op_4 (q)} =
  \gamma_{42,2} \Op_2 (p+q) + \gamma_{42,0} (q,p)\, \delta
  (p+q)\,.
\end{equation}

We now recall that the multiple product of the number operator is given by
\begin{align}
&  \op{\Op_2 (p_1) \cdots \Op_4 (q_1) \cdots \N (r_1) \cdots \N (r_m)}\nt\\
& = \N (r_1) \star \cdots \star \N (r_m) \star \op{\Op_2 (p_1) \cdots
  \Op_4 (q_1) \cdots}\,.\label{sec6-star-products}
\end{align}
This is equivalent to
\begin{align}
  &\frac{\delta}{\delta J_\N (-p)} e^{W [J_2, J_4, J_\N]}
  = \N (p) \star e^{W [J_2, J_4, J_\N]}\nt\\
&=- e^{-S} \int_q K(q) \frac{\delta}{\delta \phi (q)} \left(
  e^S  \left( \phi (q+p) + h (q+p) \frac{\delta}{\delta \phi
                                                (-q-p)} \right) e^{W [J_2, J_4, J_\N]} \right)\,.
\label{sec6-JN}
\end{align}
To show this, it suffices to derive (\ref{sec6-star-products}) from
(\ref{sec6-JN}).  Using (\ref{sec6-JN}) $m$ times, we obtain
\begin{equation}
\frac{\delta}{\delta J_\N (-p_m)} \cdots  \frac{\delta}{\delta J_\N (-p_1)} e^{W
                [J_2, J_4, J_\N]}= \N (p_1) \star \cdots \N (p_m) \star e^{W [J_2, J_4, J_\N]}\,.
\end{equation}
Differentiating this further with respect to $J_2$ and $J_4$ a number
of times and setting the sources to zero, we obtain
(\ref{sec6-star-products}).  We have thus verified (\ref{sec6-JN}).

\subsection{Constant parameters}

Let us consider a special case where the sources are constants:
\begin{subequations}
\begin{align}
  J_2 (-p)
  &= - m^2 \,\delta (p)\,,\\
  J_4 (-p)
  &= - \lambda \,\delta (p)\,,\\
  J_\N (-p)
  &= z \,\delta (p)\,.
\end{align}
\end{subequations}
We then write $W [J_2, J_4, J_\N]$ as $W (m^2, \lambda, z)$.
$S + W (m^2,\lambda, z)$ is the Wilson action of the perturbative
$\phi^4$ theory.  The parameter $z$ changes the normalization of the
field.

(\ref{sec6-W-ERG}) gives the ERG equation
\begin{equation}
  \left[ \left(2 + \beta_m (\lambda)\right) m^2 \partial_{m^2} + \beta
    (\lambda) \partial_\lambda - \gamma (\lambda) \partial_z -
    \mathcal{D} \right] e^{W (m^2, \lambda, z)} = B_0 (m^2,\lambda) \, e^{W (m^2,
    \lambda, z)}\,,
\end{equation}
where the beta functions and the anomalous dimension are given by
\begin{align}
  \beta_m (\lambda)
  &= - \sum_{k=1}^\infty \frac{1}{k!} \gamma_{4 \cdots 4 2, 2} (-
    \lambda)^k = - \frac{\lambda}{(4 \pi)^2} + \frac{\lambda^2}{(4
    \pi)^4} \frac{5}{6} + \cdots \,,\\
  \beta (\lambda)
  &= \sum_{k=2}^\infty \frac{1}{k!} \gamma_{4 \cdots 4, 4}
    (-\lambda)^k = - \frac{3 \lambda^2}{(4 \pi)^2} +
    \frac{\lambda^3}{(4 \pi)^3} \frac{17}{3} + \cdots \,,\\
  \gamma (\lambda)
  &= \sum_{k=2}^\infty \frac{1}{k!} \gamma_{4 \cdots 4, \N}
    (-\lambda)^k = \frac{\lambda^2}{(4 \pi)^2\,12}  + \cdots\,.
\end{align}

The correlation functions are given by
\begin{equation}
  \vvev{\phi (p_1) \cdots \phi (p_n)}_{m^2, \lambda, z}
  \equiv \vvev{e^{W (m^2, \lambda, z)} \phi (p_1) \cdots \phi
    (p_n)}\,.
\end{equation}
Differentiating this with respect to $z$, we obtain
\begin{align}
  \partial_z \vvev{\phi (p_1) \cdots \phi (p_n)}_{m^2, \lambda, z}
&=  \vvev{\N (0) \star e^{W (m^2, \lambda, z)} \phi (p_1) \cdots \phi
                                                                     (p_n)}\nt\\
  &= n \vvev{\phi (p_1) \cdots \phi (p_n)}_{m^2, \lambda, z}\,.
\end{align}
Hence, we obtain the $z$-dependence as
\begin{equation}
  \vvev{\phi (p_1) \cdots \phi (p_n)}_{m^2, \lambda, z}
  = e^{n z}  \vvev{\phi (p_1) \cdots \phi (p_n)}_{m^2, \lambda,
    z=0}\,.
\end{equation}

\subsection{Source for the identity operator}

We find it handy to introduce a source $J_0 (p)$  that couples to the
identity operator $\delta (p)$:
\begin{equation}
  W [J_0, J_2, J_4, J_\N] = \int_p J_0 (-p) \delta (p) + W [J_2, J_4,
  J_\N] = J_0 (0) + W[J_2. J_4, J_\N]\,.
\end{equation}
We can then rewrite the ERG equation (\ref{sec6-W-ERG}) as
\begin{align}
&  \int_p \Bigg[  \lb - p \cdot \partial_p J_0 (-p) - B_0 [J_2,
                J_4] \rb \frac{\delta}{\delta J_0 (-p)}\nt\\
  &\quad + \lb \left( - p \cdot \partial_p - 2 \right) J_2 (-p) - B_2
    (-p) \rb \frac{\delta}{\delta J_2 (-p)}\nt\\
  &\quad  + \lb \left( - p \cdot \partial_p - 4 \right) J_4 (-p) - B_4
    (-p) \rb \frac{\delta}{\delta J_4 (-p)}\nt\\
  &\quad  + \lb \left( - p \cdot \partial_p - 4 \right) J_\N (-p) - B_\N
    (-p) \rb \frac{\delta}{\delta J_\N (-p)} - \mathcal{D} \Bigg] e^{W
    [J_0, J_2, J_4, J_\N]} =   0\,.
\end{align}

\subsection{Change of parameters}

In Sec.~\ref{section-products-two} we have constructed the products of
two composite operators by solving the respective ERG differential
equations.  Some of the equations do not have unique solutions, and we
have made arbitrary choices.  For example, we have found $\op{\Op_2 (p)
  \Op_2 (q)}$ is ambiguous by a constant multiple of $\delta (p+q)$.

This ambiguity is related to our freedom of changing parameters (or
sources) as long as we preserve scale dimensions and respect
locality.  In general we can introduce the following change of parameters:
\begin{align}
  J'_0 (-p)
  &= J_0 (-p) \nt\\
  &\quad + \sum_{k=1}^\infty \frac{1}{k!} C_{0,k} \frac{1}{2}
    \int_{p_1,\cdots,p_k, q,l} J_2 (-q) J_2 (-l)  J_4 (-p_1) \cdots
    J_4 (-p_k)\, \delta \left( \sum_{i=1}^k p_i + q + l - p \right)\nt\\
  &\quad + \sum_{k=0}^\infty \frac{1}{k!} \int_{p_1,\cdots, p_k, q}
    J_2 (-q) J_4 (-p_1) \cdots J_4 (-p_k)\, C'_{0,k} (q; p_1, \cdots,
    p_k)\,
    \delta \left( \sum_{i=1}^k + q - p\right)\nt\\
  &\quad + \sum_{k=1}^\infty \frac{1}{k!} \int_{p_1, \cdots, p_k} J_4
    (-p_1) \cdots J_4 (-p_k)\, C''_{0,k} (p_1, \cdots, p_k)\, \delta
    \left( \sum_{i=1}^k p_i - p\right)\,,
\end{align}
where $C'_{0,k}, C''_{0,k}$ are quadratic, quartic in momenta,
respectively.
\begin{align}
    J'_2 (-p)
    &= J_2 (-p) + \sum_{k=1}^\infty \frac{1}{k!} C_{2,k} \int_{p_1,
      \cdots, p_k, q} J_2 (-q) J_4 (-p_1) \cdots J_4 (-p_k) \, \delta
      \left( q + \sum_{i=1}^k p_i - p \right)\nt\\
  &\quad + \sum_{k-1}^\infty \frac{1}{k!} \int_{p_1, \cdots, p_k}
    C'_{2,k} (p_1, \cdots, p_k)\, \delta
    \left( q + \sum_{i=1}^k p_i - p \right)\,,
\end{align}
where $C'_{2,k}$ are quadratic in momenta, and
\begin{align}
  J'_4 (-p)
  &= J_4 (-p) + \sum_{k=2}^\infty \frac{1}{k!} C_{4,k} 
\int_{p_1,
      \cdots, p_k} J_4 (-p_1) \cdots J_4 (-p_k) \, \delta
    \left( q + \sum_{i=2}^k p_i - p \right)\,.
\end{align}

The above change of parameters should not change the generating
functional in the sense that
\begin{equation}
  e^{W [J_2, J_4, J_\N]} = e^{W' [J'_2, J'_4, J_\N]}\,.
\end{equation}
The products of composite operators, defined as differentials of the
generating functional, change accordingly.  For example,
\begin{align}
  \op{\Op_2 (p) \Op_2 (q)}
  &\equiv \frac{\delta^2}{\delta J_2 (-p) \delta J_2 (-q)} e^{W
    [J_2, J_4, J_\N]} \Big|_{J_2=J_4=J_\N=0}\nt\\
  &= \frac{\delta^2}{\delta J'_2 (-p) \delta J'_2 (-q)} e^{W'
    [J'_2, J'_4, J_\N]}\Big|_{J'_2=J'_4=J_\N=0}+  \frac{\delta^2 J'_0 (0)}{\delta J_2 (-p) \delta J_2
    (-q)} \Big|_{J_2=J_4=J_\N=0}\nt\\
  &= \op{\Op_2 (p) \Op_2 (q)}' + C_{0,0} \delta (p+q)\,.
\end{align}

As we saw in some details in Sec.~\ref{section-multiple}, redefined
multi-operator products have different mixing coefficients.  This is
easy to understand since the mixing coefficients define the beta
functions $B_0, \cdots, B_4$ of the parameters, and the redefined
parameters obtain different beta functions $B'_0, B'_2$, and $B'_4$.
These beta functions are easier to think of in coordinate space, and
we give a general form of the beta functions and consider their
simplest form in Appendix \ref{appendix:coordinates}.

\section{Conclusions\label{section-conclusions}}

In this paper we have used the exact renormalization group (ERG)
formalism to construct the multiple products of composite operators at
the Gaussian fixed-point in $D=4$ dimensions.  We have considered only
three scalar composite operators $\Op_2, \Op_4, \N$ whose scale
dimensions in momentum space are $-2, 0, 0$, respectively.  Their
products do not generate any new operators, and the algebra is closed
in this sense.  We have shown, in Sec.~\ref{section-multiple}, that
the unintegrable short-distance singularities of their multiple
products determine the mixing coefficients under scaling.  The mixing
coefficients in turn determine the scaling properties of the sources
coupled to the three operators as given in (\ref{sec6-W-ERG}) to
(\ref{sec6-BN}).  Though these results are expected, we would like to
emphasize the ease and clarity with which they are drawn naturally out
of the ERG formalism.

On the technical side, our choice of the multiple products of the
number operator (\ref{sec5-multiple-N}) greatly simplifies the
dependence of the theory on the source $J_\N$.  The dependence is just
as what we expect naively from the change of field normalization, and
the products with the number operator do not generate any
short-distance singularity.

It would be nice to apply the ERG formalism to study the operator
algebra at a more nontrivial fixed-point such as the Wilson-Fisher
fixed-point in $2 < D < 4$ dimensions.

\appendix

\section{Calculations of the mixing coefficients\label{appendix:integrals}}

For the calculations of the mixing coefficients, we need to compute
the following integrals:
\begin{align}
  I_1
  &\equiv \int_q f(q) h (q)\,,\\
  I_2
  &\equiv \frac{d}{dp^2} \int_q f(q) F(q+p)\Big|_{p=0}\,,\\
  I_3
  &\equiv \frac{1}{2} \frac{d^2}{(dp^2)^2} \int_q f(q) G
    (q+p)\Big|_{p=0}\,,\\
  I_4
  &\equiv \int_p f(p) \int_q h(q)^2 h(p+q) + 2 \int_p f(p) h(p)
    F(p)\,.
\end{align}
All these integrals are universal, i.e., their values are independent
of the choice of a cutoff function $K (p)$ or equivalently $h(p)
\equiv \frac{1-K(p)}{p^2}$.  Let us show the universality of the first
two integrals.  Under an infinitesimal change $\delta h$ of $h$, we
find
\begin{align}
  \delta I_1
  &= \int_q \left( \delta f(q)\cdot h(q) + f(q) \cdot \delta h
    (q)\right)\nt\\
  &= \int_q \left( \left(q \cdot \partial_q + 2\right) \delta h(q)
    \cdot h (q) + f(q) \delta h(q) \right)\nt\\
  &= \int_q \delta h (q) \left( - \left( q \cdot \partial_q + 2
    \right) h (q) + f(q) \right) = 0\,.
\end{align}
For $I_2$, we need
\begin{align}
  \delta F (p)
  &= \delta \frac{1}{2} \int_q h (q) \left( h(q+p)-h(q)\right)\nt\\
  &= \int_q \delta h(q) \left( h(q+p) h (q)\right)\,.
\end{align}
Hence, we obtain
\begin{align}
  \delta \int_q f(q) F (q+p)
  &= \int_q \left( \delta f(q) \cdot F(q+p) + f(q) \delta F(q+p)
    \right)\nt\\
  &= \int_q \left[ \delta h(q) (-q \cdot \partial_q - 2) F(q+p)
    + f(q) \int_r \delta h(r) \left( h(r+q+p) - h(r) \right)
    \right)\nt\\
  &= \int_q \delta h (q) \Big[ (- q \cdot \partial_q - p \cdot
    \partial_p) F(q+p) + (p \cdot \partial_p - 2) F(q+p)\nt\\
&\qquad +    \int_r f(r) \left( h(r+q+p)-h(q)\right) \Big]\nt\\
  &= (p \cdot \partial_p - 2) \int_q \delta h(q) F(q+p)
 + \int_q \delta h(q)  \int_r f (r) \left( h(r) - h(q)\right)\,,
\end{align}
where the first integral has no $p^2$ term, and the second integral is
a constant.  Hence,
\begin{equation}
  \delta I_2 = 0\,.
\end{equation}

Since the integrals are independent of the choice of cutoff function,
we can make any appropriate choice such as
\begin{subequations}
\begin{align}
  K(p) &= e^{- p^2}\,,\\
  h(p) &= \int_0^1 ds\, e^{- s p^2} = \frac{1 - e^{-p^2}}{p^2}\,,\\
  f(p) &= \left( p \cdot \partial_p + 2\right) h(p) = 2 e^{- p^2}\,.
\end{align}
\end{subequations}
The advantage of this choice is that all the momentum integrals become
Gaussian.  The results are as follows:
\begin{align}
  I_1 &= \frac{1}{(4 \pi)^2}\,\\
  I_2 &= - \frac{1}{(4 \pi)^4} \frac{1}{6}\,\\
  I_3 &= \frac{1}{(4 \pi)^6} \frac{1}{144}\,\\
  I_4 &= \frac{1}{(4 \pi)^4}\,.
\end{align}
We only calculate $I_2$ here.  We first calculate
\begin{align}
  F (p)
  &= \frac{1}{2} \int_q h(q) \left( h (q+p) - h (q)\right)\nt\\
  &= \frac{1}{2} \int_q \int_0^1 ds\, e^{- s q^2} \int_0^1 dt\, \left(
    e^{- t (q+p)^2} - e^{- t q^2} \right)\nt\\
  &= \frac{1}{2} \frac{1}{(4 \pi)^2} \int_0^1 ds \int_0^1 dt\,
    \frac{1}{(s+t)^2} \left( e^{- \frac{st}{s+t} p^2} - 1 \right)\,.
\end{align}
This gives
\begin{equation}
  \int_q f(q) F (q+p)
  = \frac{1}{(4 \pi)^4} \int_0^1 ds \int_0^1 dt\, \frac{1}{(s+t+st)^2}
  \left( e^{- \frac{st}{s+t+st} p^2} - 1 \right)\,.
\end{equation}
Hence, we obtain
\begin{align}
  \frac{d}{dp^2} \int_q f(q) F (q+p)\Big|_{p=0}
  &= \frac{1}{(4 \pi)^4} \int_0^1 ds \int_0^1 dt\,
    \frac{- st}{(s+t+st)^3}\nt\\
  &= - \frac{1}{(4 \pi)^4} \frac{1}{6}\,.
\end{align}

\section{Asymptotic behavior of $G (p), H (p)$\label{appendix:H}} 

$G (p)$ satisfies the differential equation (\ref{sec4-G-diffeq})
\begin{equation}
  \left( p \cdot \partial_p - 2 \right) G (p) = \int_q f(q) F(q+p) + 2
  \gamma_{42,2} v_2 + \eta \,p^2\,.
\end{equation}
Hence, using (\ref{sec4-F-diffeq}) satisfied by $F$, we obtain
\begin{equation}
  \left( p \cdot \partial_p - 2 \right) \left( G (p) - 2 v_2 F (p)
  \right) = \int_q f(q) \left( F (q+p) - F (p) \right) + \eta \,
  p^2\,.
\end{equation}
This gives
\begin{equation}
  \left( p \cdot \partial_p - 2 \right) \left( G (p) - 2 v_2 F (p)
  \right) \overset{p \to \infty}{\longrightarrow} \eta\,p^2\,.
\end{equation}
Hence, we obtain the asymptotic behavior
\begin{equation}
  G (p) \overset{p \to \infty}{\longrightarrow} \eta\, p^2 \ln p + 2
  \gamma_{42,2} v_2 \ln p\,,\label{appendix-G-asymp}
\end{equation}
where we have dropped constant multiples of $p^2, 1$.
  
$H(p)$ satisfies the differential equation (\ref{sec4-H-diffeq}):
\begin{equation}
  \left( p \cdot \partial_p - 4 \right) H (p)
  = \int_q f(q) G (q+p) + v_2^2 \frac{1}{3} \gamma_{44,4} - \eta
  \int_q K(q) + 2 \eta p^2 v_2 + \gamma_{44,0} \cdot p^4\,.
\end{equation}
This gives
\begin{align}
&  \left( p \cdot \partial_p - 4 \right) \left( H(p) - 2 v_2 G (p) + 2
                v_2^2 F (p) \right)\nt\\
  &= \int_q f(q) \left( G(q+p) - G (p) \right) - 2 v_2 \int_q f(q)
    \left( F(q+p) - F(p) \right) + 2 v_2^2 \int_q f(q) h (q+p)\nt\\
  &\quad + \gamma_{44,0}\, p^4 - \eta \int_q K(q)\,.
\end{align}
Hence, we obtain
\begin{equation}
  \left( p \cdot \partial_p - 4 \right) \left( H(p) - 2 v_2 G (p) + 2
                v_2^2 F (p) \right) \overset{p \to
                \infty}{\longrightarrow}
              \gamma_{44,0} \, p^4 - \eta \int_q K(q)\,.
\end{equation}
This gives the asymptotic behavior            
\begin{equation}
  H (p) \overset{p \to \infty}{\longrightarrow} \ln p \cdot \left( \gamma_{44,0}\, p^4
    + 2 \eta \,p^2 v_2  + 2 \gamma_{42,2} v_2^2 \right)\,,
\end{equation}
where we have ignored constant multiples of $p^4, p^2, 1$.

\section{Equation-of-motion composite operators\label{appendix:eom}}

For a composite operator $\Op (p)$ of momentum $p$, we define
an equation-of-motion composite operator by
\begin{equation}
  \mathcal{E}_\Op (p) \equiv - e^{-S} \int_q K(q) \frac{\delta}{\delta
    \phi (q)} \left( \Op (q+p)\, e^S \right)\,.
\end{equation}
By definition (\ref{sec2-Op-corr}), we obtain, using functional integration by parts,
\begin{align}
  &\vvev{\mathcal{E}_\Op (p) \, \phi (p_1) \cdots \phi (p_n)}\notag\\
  &= \prod_{i=1}^n \frac{1}{K(p_i)}\, \vev{\mathcal{E}_\Op (p)\, \exp
    \left( - \int_r \frac{K(r) \left(1 - K(r)\right)}{r^2} \frac{1}{2}
    \frac{\delta^2}{\delta \phi (r) \delta \phi (-r)} \right)\, \phi
    (p_1) \cdots \phi (p_n)}_S\notag\\
  &= \prod_{i=1}^n \frac{1}{K(p_i)} \int_q K(q) \left\langle \Op (q+p) \exp
    \left( - \int_r \frac{K(r) \left(1 - K(r)\right)}{r^2} \frac{1}{2}
    \frac{\delta^2}{\delta \phi (r) \delta \phi (-r)}
    \right)\right.\notag\\
  &\left.\qquad\qquad\qquad \times \frac{\delta}{\delta
    \phi (q)} \left( \phi (p_1) \cdots \phi  (p_n) \right)\right\rangle_S\notag\\
  &= \prod_{i=1}^n \frac{1}{K(p_i)} \sum_{j=1}^n K(p_j) \left\langle
    \Op (p_j+ p) \exp \left( - \int_r \frac{K(r) \left(1 - K(r)\right)}{r^2} \frac{1}{2}
    \frac{\delta^2}{\delta \phi (r) \delta \phi (-r)}
    \right)\right.\notag\\
  &\left.\qquad\qquad\qquad \times  \phi (p_1) \cdots
    \widehat{\phi (p_j)} \cdots \phi (p_n) \right\rangle_S\notag\\
  &= \sum_{j=1}^n \vvev{\Op (p_j+p)\,  \phi (p_1) \cdots
    \widehat{\phi (p_j)} \cdots \phi (p_n) }\,,
\end{align}
where $\phi (p_j)$ under the hat is omitted.

Since the elementary field
$\phi$ is also a composite operator for the free theory\footnote{This
  is not the case for interacting theories.},  we can choose $\Op (p) =
\phi (p)$ to obtain 
\begin{equation}
  \mathcal{E}_\phi (p) = \N (p)\,,
\end{equation}
which is Eq.~(\ref{sec2-def-N}), and
\begin{equation}
  \vvev{\N (p)\, \phi (p_1) \cdots \phi (p_n)} = \sum_{i=1}^n
  \vvev{\phi (p_1) \cdots \phi (p_i+p)\cdots \phi (p_n)}\,,
\end{equation}
which is Eq.~(\ref{sec5-N-corr}).

\section{Products of three composite operators\label{appendix:three}}

In this appendix we sketch the calculations of the products of three
composite operators.

\subsection{$\op{\Op_2 \Op_2 \Op_2}$}

The counterterm $\PP_{222} (p,q,r)$ satisfies the ERG equation with no
mixing:
\begin{align}
  & \left( p \cdot \partial_p + q \cdot \partial_q + r \cdot
    \partial_r + 6 - \mathcal{D} \right) \PP_{222} (p,q,r)\nt\\
  &= \int_s f(s) \left(
    \frac{\delta \PP_{22} (p,q)}{\delta \phi (s)} \frac{\delta \Op_2
    (r)}{\delta \phi (-s)} + \frac{\delta \PP_{22} (q,r)}{\delta \phi (s)} \frac{\delta \Op_2
    (p)}{\delta \phi (-s)} + \frac{\delta \PP_{22} (r,p)}{\delta \phi (s)} \frac{\delta \Op_2
    (q)}{\delta \phi (-s)} \right)\,.
\end{align}
The solution is given by
\begin{equation}
  \PP_{222} (p,q,r) = \int_s h(s) h(s+q) h(s+q+r) =\raisebox{-0.6cm}{
  \includegraphics{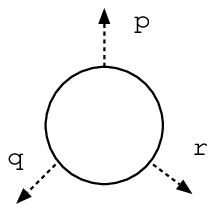}}\,.
\end{equation}

\subsection{$\op{\Op_4 \Op_2 \Op_2}$}

The ERG equation is given by
\begin{align}
  & \left( p \cdot \partial_p + q \cdot \partial_q + r \cdot
    \partial_r + 4 - \mathcal{D} \right) \PP_{422} (p,q,r)\nt\\
  &= \int_s f(s) \left( \frac{\delta \PP_{42} (p,q)}{\delta \phi (s)}
    \frac{\delta \Op_2 (r)}{\delta \phi (-s)} +  \frac{\delta \PP_{42} (p,r)}{\delta \phi (s)}
    \frac{\delta \Op_2 (q)}{\delta \phi (-s)} + \frac{\delta \PP_{22} (q,r)}{\delta \phi (s)}
    \frac{\delta \Op_4 (p)}{\delta \phi (-s)} \right)\nt\\
  &\quad + \gamma_{42,2} \left( \PP_{22} (p+q,r) + \PP_{22} (p+r,q)
    \right) + \gamma_{422,0} \,\delta (p+q+r)\,.
\end{align}
Let
\begin{equation}
  \PP_{422} (p,q,r)
  = \sum_{n=0,1,2} \frac{1}{(2n)!} \int_{p_1,\cdots,p_{2n}}
    \prod_{i=1}^{2n} \phi (p_i)\, \delta \left( \sum_i p_i - p - q -
      r\right)\, c_{422, 2n} (p,q,r; p_1, \cdots,p_{2n})\,.
\end{equation}
We obtain
\begin{equation}
  c_{422,4} (p,q,r; p_1,p_2,p_3,p_4)
  = \raisebox{-0.3cm}{\includegraphics{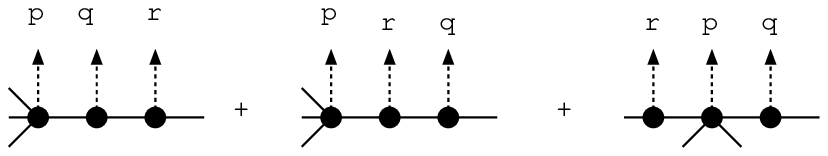}}\,,
\end{equation}
\begin{equation}
  c_{422,2} (p,q,r; p_1, p_2)
  = \raisebox{-2.8cm}{\includegraphics{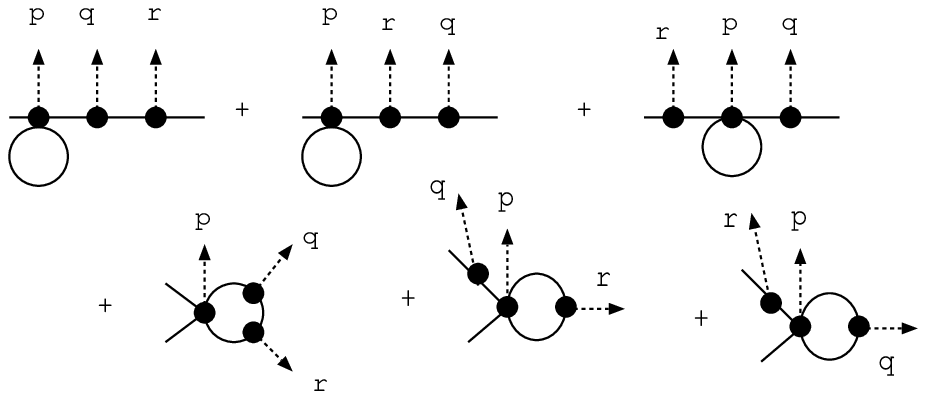}}
\end{equation}
and
\begin{align}
  c_{422,0} (p,q,r)
  &= v_2 \int_s h(s) h(s+q) h(s+q+p) + F(q) F(r)\\
  &= \raisebox{-0.8cm}{\includegraphics{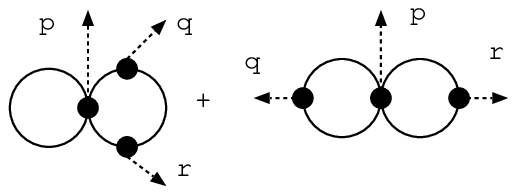}}
\end{align}
with
\begin{equation}
  \gamma_{422, 0} = 0\,.
\end{equation}
Note that the final result depends on our choice of $\PP_{22}$; if we
changed $\PP_{22} (p,q)$ by $a\, \delta (p+q)$, where $a$ is a
constant, we would obtain $\gamma_{422,0} = - 2 \gamma_{42,2}\, a$.

\subsection{$\op{\Op_4 \Op_4 \Op_2}$}

The ERG equation is given by
\begin{align}
  & \left( p \cdot \partial_p + q \cdot \partial_q + r \cdot
    \partial_r + 2 - \mathcal{D} \right) \PP_{442} (p,q,r)\nt\\
  &= \int_s f(s) \left( \frac{\delta \PP_{44} (p,q)}{\delta \phi (s)}
    \frac{\delta \Op_2 (r)}{\delta \phi (-s)} + \frac{\delta \PP_{42} (p,r)}{\delta \phi (s)}
    \frac{\delta \Op_4 (q)}{\delta \phi (-s)}  +\frac{\delta \PP_{44} (q,r)}{\delta \phi (s)}
    \frac{\delta \Op_4 (p)}{\delta \phi (-s)} \right)\nt\\
  &\quad + \gamma_{44,4} \PP_{42} (p+q,r) + \gamma_{44,\N} \PP_{\N 2}
    (p+q,r) + \gamma_{44,2} (p,q) \PP_{22} (p+q,r)\nt\\
  &\quad + \gamma_{42,2} \left( \PP_{42} (q,p+r) + \PP_{42} (p,q+r)
    \right) \nt\\
  &\quad + \gamma_{442,2} \Op_2 (p+q+r) + \gamma_{442,0} (p,q,r)
    \delta (p+q+r)\,,
\end{align}
where $\gamma_{442,0} (p,q,r)$ is quadratic.  Before we calculate, let
us remark that $\gamma_{442,2}$ depends on our choice of $\PP_{42}$.
If we changed $\PP_{42} (p,q)$ by $a \Op_2 (p+q)$, where $a$ is a
constant, $\gamma_{442,2}$ would change by $- \left(\gamma_{44,4} +
  \gamma_{42,2} \right) a$.  Similarly, $\gamma_{442,0}$ would change
if we changed $\PP_{42}, \PP_{22}$ in proportion to the identity operator.

We only consider the 1PI part of $c_{442,4}, c_{442,2}$ here.
\begin{equation}
  c_{442,4}^{\mathrm{1PI}} (p,q,r; p_1,p_2,p_3,p_4)
  = \raisebox{-1cm}{\includegraphics{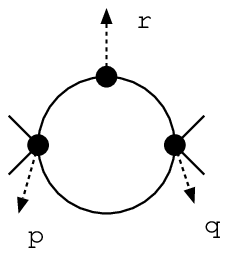}}\,,
\end{equation}
\begin{align}
  c_{442,2}^{\mathrm{1PI}} (p,q,r; p_1, p_2)
  &= \raisebox{-1cm}{\includegraphics{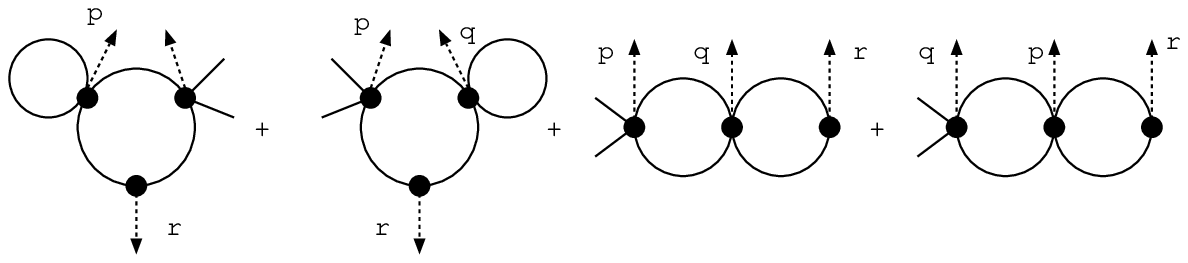}}\nt\\
  &\quad + \raisebox{-1cm}{\includegraphics{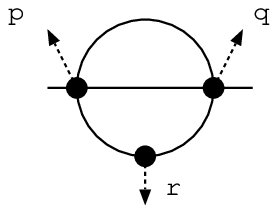}}\,,
\end{align}
where the last graph satisfies the ERG equation
\begin{align}
&  \left( p \cdot \partial_p + q \cdot \partial_q + r \cdot \partial_r
  + \sum_{i=1,2} p_i \cdot \partial_{p_i}\right)
                \raisebox{-1cm}{\includegraphics{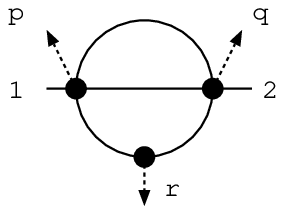}}\nt\\
    &= \int_s f(s) \Big( \int_t h (t) h (p_1-p-s-t) h(p_1-p-r-s-t)
      \nt\\
  &\qquad +
       h (s-r) F (p_1 - p - s) + h(s+r) F (p_1-p-s-r) \Big)\nt\\
&\quad + \frac{1}{3} \gamma_{44,4} F (r) - \gamma_{44,\N} +
                                                                     \frac{1}{2} \gamma_{442,2}\,.
\end{align}
$\gamma_{442,2}$ is determined so that rhs vanishes at zero momenta:
\begin{equation}
  \int_s f(s) \int_t h(t)^2 h(s+t) + 2 \int_s f(s) h(s) F (s) -
  \gamma_{44,\N} + \frac{1}{2} \gamma_{442,2} = 0\,.\label{appendix:442-2}
\end{equation}
Using the integral $I_4$ in Appendix \ref{appendix:integrals} and
$\gamma_{44,\N} = \frac{1}{(4 \pi)^4} \frac{1}{6}$, we
obtain
\begin{equation}
  \gamma_{442,2} = - \frac{1}{(4 \pi)^4} \frac{5}{3}\,.
\end{equation}

\subsection{$\op{\Op_4 \Op_4 \Op_4}$}

The ERG equation is
\begin{align}
  & \left( p \cdot \partial_p + q \cdot \partial_q + r \cdot
    \partial_r - \mathcal{D} \right) \PP_{444} (p,q,r)\nt\\
  &= \int_s f(s) \left( \frac{\delta \PP_{44} (p,q)}{\delta \phi (s)}
    \frac{\delta \Op_4 (r)}{\delta \phi (-s)} + \frac{\delta \PP_{44} (q,r)}{\delta \phi (s)}
    \frac{\delta \Op_4 (p)}{\delta \phi (-s)} +\frac{\delta \PP_{44} (r,p)}{\delta \phi (s)}
    \frac{\delta \Op_4 (q)}{\delta \phi (-s)} \right)\nt\\
  &\quad + \gamma_{44,4} \left( \PP_{44} (p+q,r) + \PP_{44} (q+r,p) +
    \PP_{44} (r+p, q)\right)\nt\\
  &\quad + \gamma_{44,\N} \left( \PP_{\N 4} (p+q,r) + \PP_{\N 4}
    (q+r,p) + \PP_{\N 4} (r+p, q) \right)\nt\\
  &\quad + \gamma_{44,2} (p,q) \PP_{24} (p+q,r) + \gamma_{44,2} (q,r)
    \PP_{24} (q+r,p) + \gamma_{44,2} (r,p) \PP_{24} (r+p, q) \nt\\
  &\quad +\gamma_{444,4} \Op_4 (p+q+r) + \gamma_{444,\N} \N (p+q+r)
    \nt\\
  &\quad +
    \gamma_{444,2} (p,q,r) \Op_2 (p+q+r) + \gamma_{444,0} (p,q,r)\,,
\end{align}
where $\gamma_{444,2}$ is quadratic, and $\gamma_{444,0}$ quartic in
momenta.

Here we only compute part of the 1PI part $c_{444,4}^{\mathrm{1PI}}
(p,q,r; p_1, p_2, p_3, p_4)$ satisfying the ERG equation
\begin{align}
  &\left( p \cdot \partial_p + \cdots + \sum_{i=1}^4 p_i \cdot
    \partial_{p_i} \right) \raisebox{-1cm}{\includegraphics{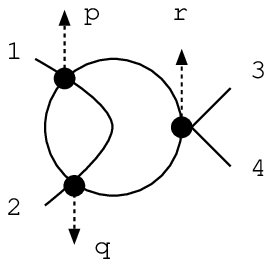}}\nt\\
  &= \int_s f(s) \Big( \int_t h(t) h (p_1 - p - s - t) h (p_1+p_3+p_4
    -p-s-t-r) \nt\\
  &\qquad + F (p_1 - p - s) h (s + p_3 + p_4 - r) + F (p_2 - q -
    s) h (s + p_3 + p+4 - r) \Big)\nt\\
  &\quad + \frac{1}{3} \gamma_{44,4} F (p_1+p_2-p-q) + \frac{1}{12}
    \left( - 4 \gamma_{44,\N} + \frac{1}{3} \gamma_{444,4} \right)\,.
\end{align}
This is the same equation (and graph) as (\ref{appendix:442-2}).
$\gamma_{444,4}$ is determined so that the rhs vanishes at zero
momenta:
\begin{equation}
   \frac{1}{12}  \left( 4 \gamma_{44,\N} - \frac{1}{3} \gamma_{444,4}
   \right)
   = \int_s f(s) \int_t h(t) h(s+t)^2 + 2 \int_s f(s) h(s) F(s) =
   \frac{1}{(4 \pi)^4}\,.
 \end{equation}
 Hence, using $\gamma_{44, \N} = \frac{1}{(4\pi)^4} \frac{1}{6}$, we obtain
 \begin{equation}
   \gamma_{444,4} = - \frac{34}{(4 \pi)^4} \,.
 \end{equation}
It turns out that even if we changed $\PP_{44} (p,q)$ by $a\, \Op_4 (p+q)$,
$\gamma_{444,4}$ would remain intact.  (If we substitute
$\sigma_{44,4} = a$ into (\ref{sec3-new-mixing}), we get zero.)
 
\section{In coordinate space\label{appendix:coordinates}}

Following \cite{Jack:1990eb, Osborn:1991gm, Osborn:1991mk}, we
consider space dependent parameters which are simply the Fourier
transforms of the momentum dependent parameters:
\begin{subequations}
\begin{align}
   \lambda (x) &= - \int_p e^{- i p x} J_4 (-p)\,, \\
   m^2 (x) &= - \int_p e^{- i p x} J_2 (-p)\,,\\
   g (x) &= - \int_p e^{- i p x} J_0 (-p)\,.
\end{align}
\end{subequations}
The RG equations of the parameters are given by
\begin{subequations}
 \begin{align}
   \frac{d}{dt} \lambda (x)
   &= x_\mu \cdot \partial_\mu \lambda (x) + B_4 (x)\,,\\
   \frac{d}{dt} m^2 (x)
   &= \left(x_\mu \cdot \partial_\mu + 2 \right) m^2 (x) + B_2 (x)\,,\\
   \frac{d}{dt} g (x)
   &= \left(x_\mu \cdot \partial_\mu + 4\right) g(x) + B_0 (x)\,,
 \end{align}
 \end{subequations}
 where
 \begin{subequations}
 \begin{align}
   B_4 (x)
   &\equiv \int_p e^{- i p x} B_4 (-p)= \beta \left(\lambda (x)\right)\,,\\
   B_2 (x)
   &\equiv \int_p e^{- i p x} B_2 (-p)\nt\\
   &= \beta_m (\lambda (x)) m^2 (x) + \beta_{m 1} (\lambda (x))
     \partial^2 \lambda (x) + \beta_{m 2} (\lambda (x)) \frac{1}{2}
     \partial_\mu \lambda (x) \partial_\mu \lambda (x)\,,\\
   B_0 (x)
   &\equiv \int_p e^{- i p x} B_0 (-p)\nt\\
   &= \beta_{0 m} (\lambda (x)) \frac{1}{2} \left( m^2 (x) \right)^2
     + \beta_{0 m 1} (\lambda (x)) \partial^2 \lambda (x) + \beta_{0 m
     2} (\lambda (x)) \frac{1}{2} \partial_\mu \lambda (x)
     \partial_\mu \lambda (x)\nt\\
   &\quad + \beta_{0 1} (\lambda (x)) (\partial^2)^2 \lambda (x) +
     \beta_{0 2} (\lambda (x)) \frac{1}{2} \partial^2 \lambda (x)
     \partial^2 \lambda (x) + \beta_{0 3} (\lambda (x)) \frac{1}{2}
     \partial_\mu \partial_\nu \lambda (x) 
     \partial_\mu \partial_\nu \lambda (x)\,.
 \end{align}
\end{subequations}
For $B_0 (x)$, we have given the most general form up to total
derivatives.

We can introduce a new parameter
\begin{equation}
  \lambda' (x) = \lambda (x) + \mathrm{O} \left( \lambda (x)^2 \right)
\end{equation}
so that its beta function is two-loop exact:
\begin{equation}
  B'_4 (x) = - \frac{3 \lambda' (x)^2}{(4 \pi)^2} + \frac{\lambda'
    (x)^3}{(4 \pi)^3} \frac{17}{3}\,.
\end{equation}
Similarly, we can introduce $m^{2 \prime} (x)$ so that
\begin{equation}
  B'_2 (x) = - m^{2\prime} (x) \frac{\lambda' (x)}{(4 \pi)^2} +
  \frac{1}{(4 \pi)^4} \frac{1}{6}
   \partial_\mu \lambda' (x) \partial_\mu \lambda' (x)
\end{equation}
exactly.  We should also be able to introduce $g' (x)$ with a simple $B'_0
(x)$.  With such a choice of parameters, we have no higher order
mixing coefficients.  This is possible only if we make a judicious
choice of solutions to the ERG differential equations.  Suppose we
have constructed products of up to $k$ operators.  The ERG equations
leave
\begin{equation}
  \op{\Op_{i_1} (p_1) \cdots \Op_{i_k} (p_k)}
\end{equation}
ambiguous up to
\begin{equation}
  \sigma_{i_1 \cdots i_k, j} (p_1, \cdots, p_k) \Op_j (p_1 + \cdots +
  p_k)
\end{equation}
where $\sigma$ is a polynomial of degree
$- \sum_{l=1}^k y_{i_l} + y_j$.  We choose
$\sigma_{i_1 \cdots i_k, j}$ so that the mixing coefficients of the
$(k+1)$-operator products cancel.  Only with this contrived choice of
lower order products, we can eliminate the mixing coefficients.

\begin{acknowledgments}
  I would like to thank Dr. Carlo Pagani for his help and interest in
  this work and for informing me of refs. \cite{Jack:1990eb,
    Osborn:1991gm, Osborn:1991mk, Baume:2014rla}.
  
\end{acknowledgments}

\bibliography{paper}

\begin{thebibliography}{13}%
\makeatletter
\providecommand \@ifxundefined [1]{%
 \@ifx{#1\undefined}
}%
\providecommand \@ifnum [1]{%
 \ifnum #1\expandafter \@firstoftwo
 \else \expandafter \@secondoftwo
 \fi
}%
\providecommand \@ifx [1]{%
 \ifx #1\expandafter \@firstoftwo
 \else \expandafter \@secondoftwo
 \fi
}%
\providecommand \natexlab [1]{#1}%
\providecommand \enquote  [1]{``#1''}%
\providecommand \bibnamefont  [1]{#1}%
\providecommand \bibfnamefont [1]{#1}%
\providecommand \citenamefont [1]{#1}%
\providecommand \href@noop [0]{\@secondoftwo}%
\providecommand \href [0]{\begingroup \@sanitize@url \@href}%
\providecommand \@href[1]{\@@startlink{#1}\@@href}%
\providecommand \@@href[1]{\endgroup#1\@@endlink}%
\providecommand \@sanitize@url [0]{\catcode `\\12\catcode `\$12\catcode
  `\&12\catcode `\#12\catcode `\^12\catcode `\_12\catcode `\%12\relax}%
\providecommand \@@startlink[1]{}%
\providecommand \@@endlink[0]{}%
\providecommand \url  [0]{\begingroup\@sanitize@url \@url }%
\providecommand \@url [1]{\endgroup\@href {#1}{\urlprefix }}%
\providecommand \urlprefix  [0]{URL }%
\providecommand \Eprint [0]{\href }%
\providecommand \doibase [0]{http://dx.doi.org/}%
\providecommand \selectlanguage [0]{\@gobble}%
\providecommand \bibinfo  [0]{\@secondoftwo}%
\providecommand \bibfield  [0]{\@secondoftwo}%
\providecommand \translation [1]{[#1]}%
\providecommand \BibitemOpen [0]{}%
\providecommand \bibitemStop [0]{}%
\providecommand \bibitemNoStop [0]{.\EOS\space}%
\providecommand \EOS [0]{\spacefactor3000\relax}%
\providecommand \BibitemShut  [1]{\csname bibitem#1\endcsname}%
\let\auto@bib@innerbib\@empty
\bibitem [{\citenamefont {Wilson}\ and\ \citenamefont
  {Kogut}(1974)}]{Wilson:1973jj}%
  \BibitemOpen
  \bibfield  {author} {\bibinfo {author} {\bibfnamefont {K.~G.}\ \bibnamefont
  {Wilson}}\ and\ \bibinfo {author} {\bibfnamefont {J.~B.}\ \bibnamefont
  {Kogut}},\ }\bibfield  {title} {\enquote {\bibinfo {title} {{The
  Renormalization group and the epsilon expansion}},}\ }\href {\doibase
  10.1016/0370-1573(74)90023-4} {\bibfield  {journal} {\bibinfo  {journal}
  {Phys. Rept.}\ }\textbf {\bibinfo {volume} {12}},\ \bibinfo {pages} {75--199}
  (\bibinfo {year} {1974})}\BibitemShut {NoStop}%
\bibitem [{\citenamefont {Cardy}(1996)}]{Cardy:1996xt}%
  \BibitemOpen
  \bibfield  {author} {\bibinfo {author} {\bibfnamefont {J.~L.}\ \bibnamefont
  {Cardy}},\ }\href {https://books.google.de/books?id=g5hfPgAACAAJ} {\emph
  {\bibinfo {title} {{Scaling and renormalization in statistical physics}}}},\
  Cambridge lecture notes in physics\ (\bibinfo  {publisher} {Cambridge
  University Press},\ \bibinfo {year} {1996})\BibitemShut {NoStop}%
\bibitem [{\citenamefont {Pagani}\ and\ \citenamefont
  {Sonoda}(2018{\natexlab{a}})}]{Pagani:2017gnd}%
  \BibitemOpen
  \bibfield  {author} {\bibinfo {author} {\bibfnamefont {C.}~\bibnamefont
  {Pagani}}\ and\ \bibinfo {author} {\bibfnamefont {H.}~\bibnamefont
  {Sonoda}},\ }\bibfield  {title} {\enquote {\bibinfo {title} {{Geometry of the
  theory space in the exact renormalization group formalism}},}\ }\href
  {\doibase 10.1103/PhysRevD.97.025015} {\bibfield  {journal} {\bibinfo
  {journal} {Phys. Rev. D}\ }\textbf {\bibinfo {volume} {97}},\ \bibinfo
  {pages} {025015} (\bibinfo {year} {2018}{\natexlab{a}})},\ \Eprint
  {http://arxiv.org/abs/1710.10409} {arXiv:1710.10409 [hep-th]} \BibitemShut
  {NoStop}%
\bibitem [{\citenamefont {'t~Hooft}(1973)}]{tHooft:1973mfk}%
  \BibitemOpen
  \bibfield  {author} {\bibinfo {author} {\bibfnamefont {G.}~\bibnamefont
  {'t~Hooft}},\ }\bibfield  {title} {\enquote {\bibinfo {title} {{Dimensional
  regularization and the renormalization group}},}\ }\href {\doibase
  10.1016/0550-3213(73)90376-3} {\bibfield  {journal} {\bibinfo  {journal}
  {Nucl. Phys. B}\ }\textbf {\bibinfo {volume} {61}},\ \bibinfo {pages}
  {455--468} (\bibinfo {year} {1973})}\BibitemShut {NoStop}%
\bibitem [{\citenamefont {Rosten}(2012)}]{Rosten:2010vm}%
  \BibitemOpen
  \bibfield  {author} {\bibinfo {author} {\bibfnamefont {O.~J.}\ \bibnamefont
  {Rosten}},\ }\bibfield  {title} {\enquote {\bibinfo {title} {{Fundamentals of
  the Exact Renormalization Group}},}\ }\href {\doibase
  10.1016/j.physrep.2011.12.003} {\bibfield  {journal} {\bibinfo  {journal}
  {Phys. Rept.}\ }\textbf {\bibinfo {volume} {511}},\ \bibinfo {pages}
  {177--272} (\bibinfo {year} {2012})},\ \Eprint
  {http://arxiv.org/abs/1003.1366} {arXiv:1003.1366 [hep-th]} \BibitemShut
  {NoStop}%
\bibitem [{\citenamefont {Igarashi}\ \emph {et~al.}(2010)\citenamefont
  {Igarashi}, \citenamefont {Itoh},\ and\ \citenamefont
  {Sonoda}}]{Igarashi:2009tj}%
  \BibitemOpen
  \bibfield  {author} {\bibinfo {author} {\bibfnamefont {Y.}~\bibnamefont
  {Igarashi}}, \bibinfo {author} {\bibfnamefont {K.}~\bibnamefont {Itoh}}, \
  and\ \bibinfo {author} {\bibfnamefont {H.}~\bibnamefont {Sonoda}},\
  }\bibfield  {title} {\enquote {\bibinfo {title} {{Realization of Symmetry in
  the ERG Approach to Quantum Field Theory}},}\ }\href {\doibase
  10.1143/PTPS.181.1} {\bibfield  {journal} {\bibinfo  {journal} {Prog. Theor.
  Phys. Suppl.}\ }\textbf {\bibinfo {volume} {181}},\ \bibinfo {pages} {1--166}
  (\bibinfo {year} {2010})},\ \Eprint {http://arxiv.org/abs/0909.0327}
  {arXiv:0909.0327 [hep-th]} \BibitemShut {NoStop}%
\bibitem [{\citenamefont {Sonoda}(2015)}]{Sonoda:2015bla}%
  \BibitemOpen
  \bibfield  {author} {\bibinfo {author} {\bibfnamefont {H.}~\bibnamefont
  {Sonoda}},\ }\bibfield  {title} {\enquote {\bibinfo {title} {{Equivalence of
  Wilson Actions}},}\ }\href {\doibase 10.1093/ptep/ptv130} {\bibfield
  {journal} {\bibinfo  {journal} {PTEP}\ }\textbf {\bibinfo {volume} {2015}},\
  \bibinfo {pages} {103B01} (\bibinfo {year} {2015})},\ \Eprint
  {http://arxiv.org/abs/1503.08578} {arXiv:1503.08578 [hep-th]} \BibitemShut
  {NoStop}%
\bibitem [{\citenamefont {Jack}\ and\ \citenamefont
  {Osborn}(1990)}]{Jack:1990eb}%
  \BibitemOpen
  \bibfield  {author} {\bibinfo {author} {\bibfnamefont {I.}~\bibnamefont
  {Jack}}\ and\ \bibinfo {author} {\bibfnamefont {H.}~\bibnamefont {Osborn}},\
  }\bibfield  {title} {\enquote {\bibinfo {title} {{Analogs for the $c$ Theorem
  for Four-dimensional Renormalizable Field Theories}},}\ }\href {\doibase
  10.1016/0550-3213(90)90584-Z} {\bibfield  {journal} {\bibinfo  {journal}
  {Nucl. Phys. B}\ }\textbf {\bibinfo {volume} {343}},\ \bibinfo {pages}
  {647--688} (\bibinfo {year} {1990})}\BibitemShut {NoStop}%
\bibitem [{\citenamefont {Osborn}(1991{\natexlab{a}})}]{Osborn:1991gm}%
  \BibitemOpen
  \bibfield  {author} {\bibinfo {author} {\bibfnamefont {H.}~\bibnamefont
  {Osborn}},\ }\bibfield  {title} {\enquote {\bibinfo {title} {{Weyl
  consistency conditions and a local renormalization group equation for general
  renormalizable field theories}},}\ }\href {\doibase
  10.1016/0550-3213(91)80030-P} {\bibfield  {journal} {\bibinfo  {journal}
  {Nucl. Phys. B}\ }\textbf {\bibinfo {volume} {363}},\ \bibinfo {pages}
  {486--526} (\bibinfo {year} {1991}{\natexlab{a}})}\BibitemShut {NoStop}%
\bibitem [{\citenamefont {Osborn}(1991{\natexlab{b}})}]{Osborn:1991mk}%
  \BibitemOpen
  \bibfield  {author} {\bibinfo {author} {\bibfnamefont {H.}~\bibnamefont
  {Osborn}},\ }\bibfield  {title} {\enquote {\bibinfo {title} {{Local
  renormalization group equations in quantum field theory}},}\ }in\ \href@noop
  {} {\emph {\bibinfo {booktitle} {{2nd JINR Conference on Renormalization
  Group}}}}\ (\bibinfo {year} {1991})\ pp.\ \bibinfo {pages}
  {128--138}\BibitemShut {NoStop}%
\bibitem [{\citenamefont {Baume}\ \emph {et~al.}(2014)\citenamefont {Baume},
  \citenamefont {Keren-Zur}, \citenamefont {Rattazzi},\ and\ \citenamefont
  {Vitale}}]{Baume:2014rla}%
  \BibitemOpen
  \bibfield  {author} {\bibinfo {author} {\bibfnamefont {F.}~\bibnamefont
  {Baume}}, \bibinfo {author} {\bibfnamefont {B.}~\bibnamefont {Keren-Zur}},
  \bibinfo {author} {\bibfnamefont {R.}~\bibnamefont {Rattazzi}}, \ and\
  \bibinfo {author} {\bibfnamefont {L.}~\bibnamefont {Vitale}},\ }\bibfield
  {title} {\enquote {\bibinfo {title} {{The local Callan-Symanzik equation:
  structure and applications}},}\ }\href {\doibase 10.1007/JHEP08(2014)152}
  {\bibfield  {journal} {\bibinfo  {journal} {JHEP}\ }\textbf {\bibinfo
  {volume} {08}},\ \bibinfo {pages} {152} (\bibinfo {year} {2014})},\ \Eprint
  {http://arxiv.org/abs/1401.5983} {arXiv:1401.5983 [hep-th]} \BibitemShut
  {NoStop}%
\bibitem [{\citenamefont {Pagani}\ and\ \citenamefont
  {Sonoda}(2018{\natexlab{b}})}]{Pagani:2017tdr}%
  \BibitemOpen
  \bibfield  {author} {\bibinfo {author} {\bibfnamefont {C.}~\bibnamefont
  {Pagani}}\ and\ \bibinfo {author} {\bibfnamefont {H.}~\bibnamefont
  {Sonoda}},\ }\bibfield  {title} {\enquote {\bibinfo {title} {{Products of
  composite operators in the exact renormalization group formalism}},}\ }\href
  {\doibase 10.1093/ptep/ptx189} {\bibfield  {journal} {\bibinfo  {journal}
  {PTEP}\ }\textbf {\bibinfo {volume} {2018}},\ \bibinfo {pages} {023B02}
  (\bibinfo {year} {2018}{\natexlab{b}})},\ \Eprint
  {http://arxiv.org/abs/1707.09138} {arXiv:1707.09138 [hep-th]} \BibitemShut
  {NoStop}%
\bibitem [{\citenamefont {Pagani}\ and\ \citenamefont
  {Sonoda}(2020)}]{Pagani:2020ejb}%
  \BibitemOpen
  \bibfield  {author} {\bibinfo {author} {\bibfnamefont {C.}~\bibnamefont
  {Pagani}}\ and\ \bibinfo {author} {\bibfnamefont {H.}~\bibnamefont
  {Sonoda}},\ }\bibfield  {title} {\enquote {\bibinfo {title} {{Operator
  product expansion coefficients in the exact renormalization group
  formalism}},}\ }\href {\doibase 10.1103/PhysRevD.101.105007} {\bibfield
  {journal} {\bibinfo  {journal} {Phys. Rev. D}\ }\textbf {\bibinfo {volume}
  {101}},\ \bibinfo {pages} {105007} (\bibinfo {year} {2020})},\ \Eprint
  {http://arxiv.org/abs/2001.07015} {arXiv:2001.07015 [hep-th]} \BibitemShut
  {NoStop}%
\end{thebibliography}%

\end{document}